\documentclass[english]{amsart}
\usepackage[T1]{fontenc}
\usepackage{color}
\usepackage{prettyref}
\usepackage{units}
\usepackage{textcomp}
\usepackage{mathrsfs}
\usepackage{amsthm}
\usepackage{amssymb}
\usepackage{stackrel}
\usepackage{graphicx}
\usepackage{rotfloat}

\makeatletter

\newcommand*\LyXZeroWidthSpace{\hspace{0pt}}
\providecommand{\tabularnewline}{\\}

\numberwithin{equation}{section}
\numberwithin{figure}{section}
\theoremstyle{plain}
\newtheorem*{assumption*}{\protect\assumptionname}
\theoremstyle{definition}
\newtheorem*{condition*}{\protect\conditionname}
\theoremstyle{remark}
\newtheorem*{note*}{\protect\notename}
\theoremstyle{plain}
\newtheorem{thm}{\protect\theoremname}
\theoremstyle{remark}
\newtheorem{rem}[thm]{\protect\remarkname}
\theoremstyle{definition}
\newtheorem*{sol*}{\protect\solutionname}

\usepackage [latin1]{inputenc}

\makeatother

\usepackage{babel}
\providecommand{\assumptionname}{Assumption}
\providecommand{\conditionname}{Condition}
\providecommand{\notename}{Note}
\providecommand{\remarkname}{Remark}
\providecommand{\solutionname}{Solution}
\providecommand{\theoremname}{Theorem}

\begin{document}
\title{Dynamic Analysis of Flexible Stepping Frames for Earthquakes}
\author{Arzhang Alimoradi and James L. Beck}
\date{April 25, 2025\\
{*} Arzhang Alimoradi: NIRAS A/S, Allerød, Denmark\\
{*}{*} James L. Beck: California Institute of Technology, Pasadena,
CA, U.S.A}
\address{First Author: NIRAS A/S\\
Sortemosevej 19\\
3450 Allerød\\
Denmark\\
Second Author: California Institute of Technology\\
Mail Code 9-94 \\
Pasadena, CA 91125\\
U.S.A.}
\email{arzhang.alimoradi@gmail.com}
\keywords{Instability of Stepping Frames, Aperiodic (Chaotic) Rocking Response,
Moving Resonance, Seismic Analysis, Bridge Structures.}
\thanks{Acknowledgment: Benno Rumpf, Ph.D., SMU Department of Mathematics.
Daniel Trugman of Nevada Seismological Laboratory for discussion of
precariously balanced rocks in seismogenic regions. Trantech Engineering,
LLC. The Strong-motion Virtual Data Center and PEER Center at University
of California, Berkeley. }
\subjclass[2000]{Primary: 74H45, Secondary: 70K30, 70K50, 37N99, 74H99}
\begin{abstract}
This paper investigates the nonlinear dynamics of stepping flexible
frames under seismic excitation. The conventional iterative method
of solution of peak quasi-dynamic displacement of stepping frames
is not guaranteed to converge. To address this limitation, we present
closed-form solutions and stability criteria for displacement response
of stepping flexible frames. Bifurcation of displacements in response
of such systems is next studied through the extension of dynamics
of stepping rigid bodies. An approximate analytical expression is
presented to account for the effects of moving resonance under earthquake
ground motions. The closed-form solutions for displacement demand
can be readily adjusted to incorporate the influence of moving resonance
on the quasi-dynamic response of stepping oscillators. While the quasi-dynamic
method of analysis may be useful in the early stages of design, numerical
integration of the nonlinear system of differential equations of motion
is recommended for the solution of dynamic response in such applications.
Implications for formal limit-state analysis of stepping response
are discussed, accompanied by several examples demonstrating the procedures. 

\tableofcontents{}
\end{abstract}

\maketitle

\section{Introduction}

Dynamic response of rocking bodies (also called stepping response)\footnote{Rocking and stepping are semantically different but they are used
interchangeably in this paper. Rocking is often used to describe rigid
body rotations about their base whereas stepping refers to rotations
or displacements of rigid or flexible frames about their base.} has re-emerged as an important topic of study due to both its rich
analytical intricacies and its wide range of applications. Theoretically,
dynamics of rocking bodies involves nonlinear behavior and phenomena
such as contact mechanics, sliding motion, friction, free-flight,
and various types of bifurcations \cite{Hogan-1,Hogan-3,Srinivasan}.
Pragmatically, understanding rocking body dynamics is required to
safely design certain engineering systems and to analyze some natural
systems. Examples are the stability of large-capacity gravity energy
storage structures \cite{Rosakis-1}, graphite blocks in nuclear reactor
cores \cite{Olsen}, protection of museum artifacts \cite{Calio},
preservation of historical monuments and minarets during earthquakes
\cite{Zhong}, study of precariously balanced rocks to constrain intensity
of past ground shaking in seismological field studies \cite{Anoosheh-1,Wittich-1,Anoosheh-2,Trugman-1,Hall-1},
and accelerated bridge construction techniques \cite{Vassiliou-3,Mashal,Piras},
among other applications. 

Observations of survival of tall slender structures during the May
1960 Chilean earthquake resulted in a seminal paper by Housner \cite{Housner-2}
that enabled future investigations through establishing the basic
theory and solutions for an ideal rocking rigid block. This rather
simple model was later shown to exhibit some very complicated dynamics
\cite{Hogan-2} despite the fact that stepping responses of natural
and constructed structures have been a subject of interest since antiquity
\cite{Zhong}. The next major development, and a first modern application
of stepping response in engineered structures, is the design and construction
of the S. Rangitikei Viaduct in the 1970s, an elegant structure across
the Rangitikei river in the North Island of New Zealand whose longest
span of 56 m is supported by 76 m tall piers \cite{Beck-1}. The theory
of rocking rigid blocks was extended to that of stepping flexible
A-frames (the original design concept for the Rangitikei Viaduct)
with the equations of motion and numerical solution of response developed
for the unstepped and stepped phases of motion to establish feasibility
of controlled rocking as a means of safe economic design for tall
bridge piers in seismic regions \cite{Beck-1}. 

More recently, design of structures with minimal damage after natural
disasters (a design philosophy termed ``resiliency'' \cite{Mar,Krawinkler-1,Panian})
has led to the successful implementation of controlled rocking of
large structural frames, making the study of stepping response more
relevant. In the context of modern earthquake engineering, it is prudent,
and sometimes necessary, to allow pier rocking despite the common
design tradition of avoiding such instabilities, that is, temporary
loss of stability under controlled conditions can be advantageous.
The improved seismic performance comes from natural period elongation
and increased damping of stepping structures that limit design forces,
damage, and post-elastic deformations, while being economical and
easy to construct. These attributes have rendered the concept of stepping
response as the original form of seismic isolation, a nod of appreciation
to this seasoned technology. 

In spite of its rich background, modern seismic design standards and
specifications are only in the early stages of being adopted for controlled
rocking \cite{AASHTO-1,AASHTO-2}. Our first objective in this paper
is to analyze numerical stability and convergence of these early seismic
design procedures using one-dimensional iterated maps. We propose
closed-form solutions of peak design displacement and present its
linear stability. This is described under Part 1. 

Since future developments will likely evolve around probabilistic
performance-based nonlinear dynamic response assessment under critical
forcing functions in order to establish acceptable performance criteria,
our second objective is to extend the work on the dynamics of rigid
blocks to flexible portal frames and to investigate nonlinear dynamics
of such systems subject to strong ground motion excitation. We will
also address, in Part 2, the existence of deterministic chaos and
its role in stability analysis of frames, whether dynamic stability
after static instability is plausible, and the existence of various
bifurcations. 

The paper is concluded with a discussion of the novelties of stepping
flexible frame dynamics and areas of future research needed. 

\subsection{Evidence, Relevance, and Justification}

\begin{figure}
\begin{centering}
\includegraphics[scale=0.7]{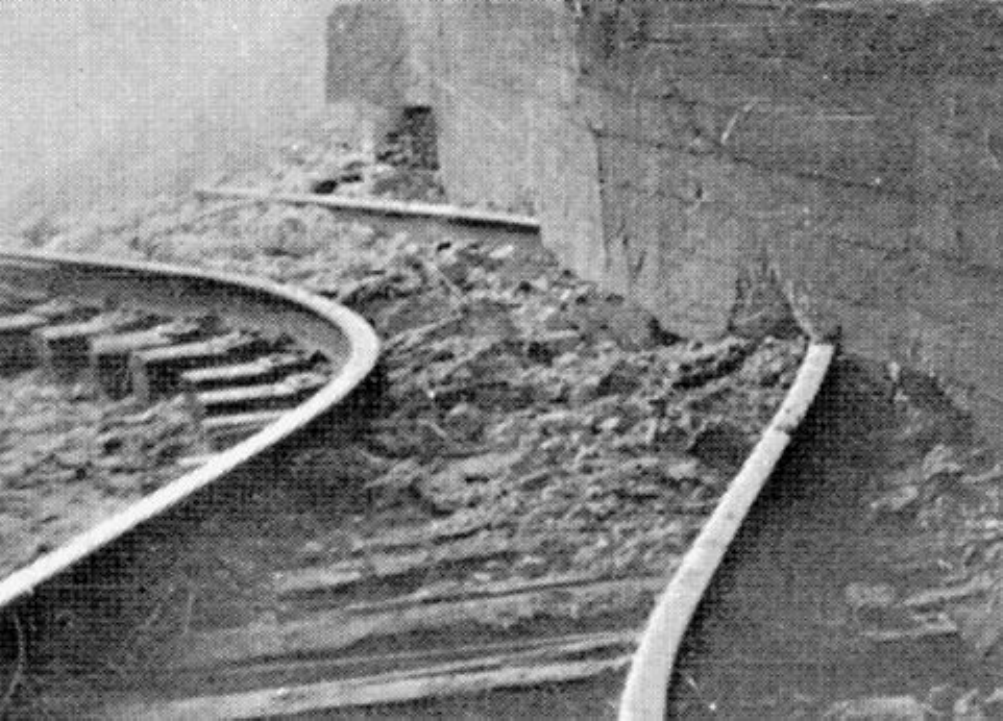}
\par\end{centering}
\caption{\label{fig:Evidence-of-rocking}Evidence of rocking and sliding motion
of a rigid wall inside Tunnel No. 3 during July 21, 1952 Kern County,
California, earthquake \cite{Caltech-1}.}

\end{figure}

The existence of rocking motion in seismic response may have been
somewhat overlooked in early investigations of post-earthquake behavior
of structures \cite{Caltech-1,Housner-1} because the primary source
of damage at the time (and still largely, today) was attributed to
horizontal forces and displacements induced by ground motion. This
is despite some early observations of rocking motion in seismic response.
Shown in Fig. \ref{fig:Evidence-of-rocking} are the bent rails of
a section of the Southern Pacific Railroad during the July 21, 1952
Kern County Mw 7.5 earthquake in Southern California showing an ``unusual
phenomenon'' of a continuous rail ``underneath the tunnel wall,
indicating that the wall {[}had{]} lifted up enough for the rail to
slide underneath.'' \cite{Caltech-1}. As a matter of fact, contributions
of foundation flexibility and rotations to dynamics of structures
has been observed in forced vibration tests \cite{Hjelmstad-1} and
system identification from earthquake response data \cite{Todorovska-1}. 

\begin{figure}
\begin{centering}
\includegraphics[scale=0.25]{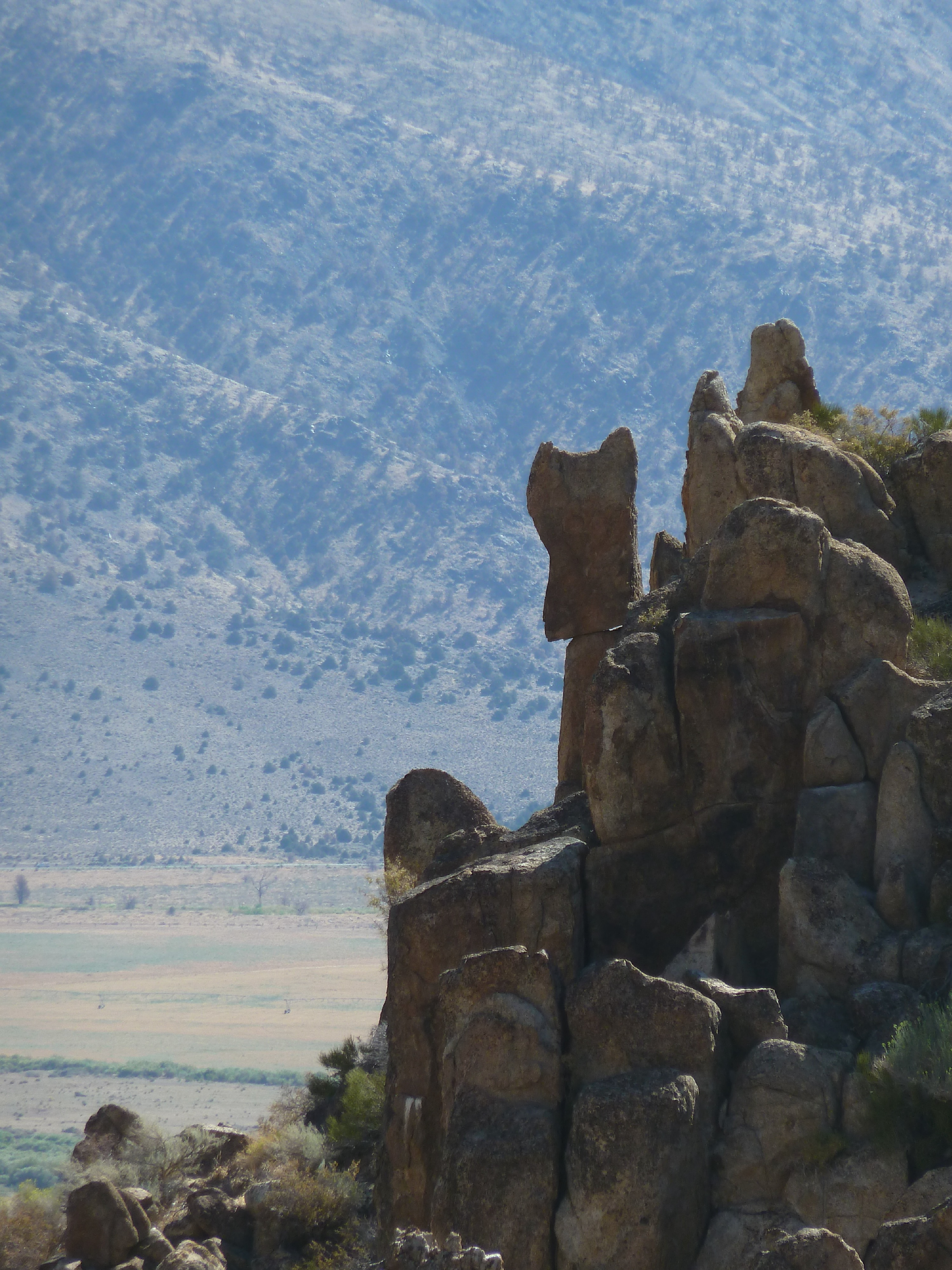}
\par\end{centering}
\caption{\label{fig:Example-of-precarious}Example of precarious rocks in Meadowcliff
Canyon near the California-Nevada border (Image Courtesy of D. Trugman).}

\end{figure}

An interesting problem in natural settings is the observation of precariously
balanced rocks (PBRs) in historically seismically active regions of
the world and whether their existence and age can tell us something
about the largest earthquakes that they have experienced. A case of
a well-studied PBR is shown in Fig. \ref{fig:Example-of-precarious}.
In several studies, to empirically constrain estimates of ground shaking
(e.g., toppling peak ground acceleration), the rocks and their conditions
have been modeled using Housner's formulation assuming a homogeneous
unattached rigid body prior to and after experiencing seismic shaking
\cite{Anoosheh-1,Housner-2,Anoosheh-2,Trugman-1}. It has also been
concluded through laboratory shaking table tests and post-earthquake
field observations that rigid body dynamics predictions generally
agree well with field conditions but improvements in modeling rock
conditions and the shaking characteristics are desirable.

\begin{figure}
\begin{centering}
\includegraphics[scale=0.125]{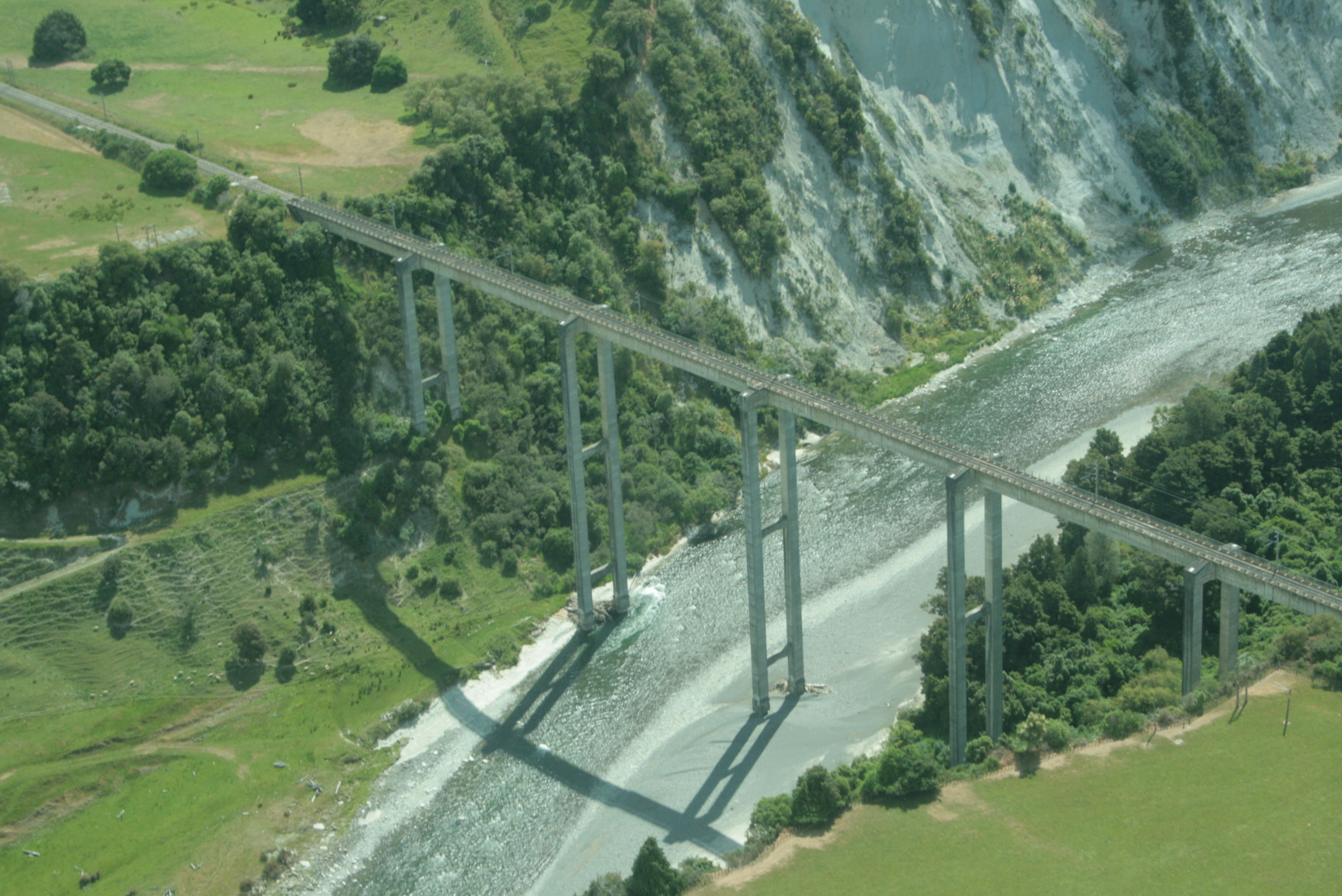}
\par\end{centering}
\begin{centering}
\includegraphics[scale=0.52]{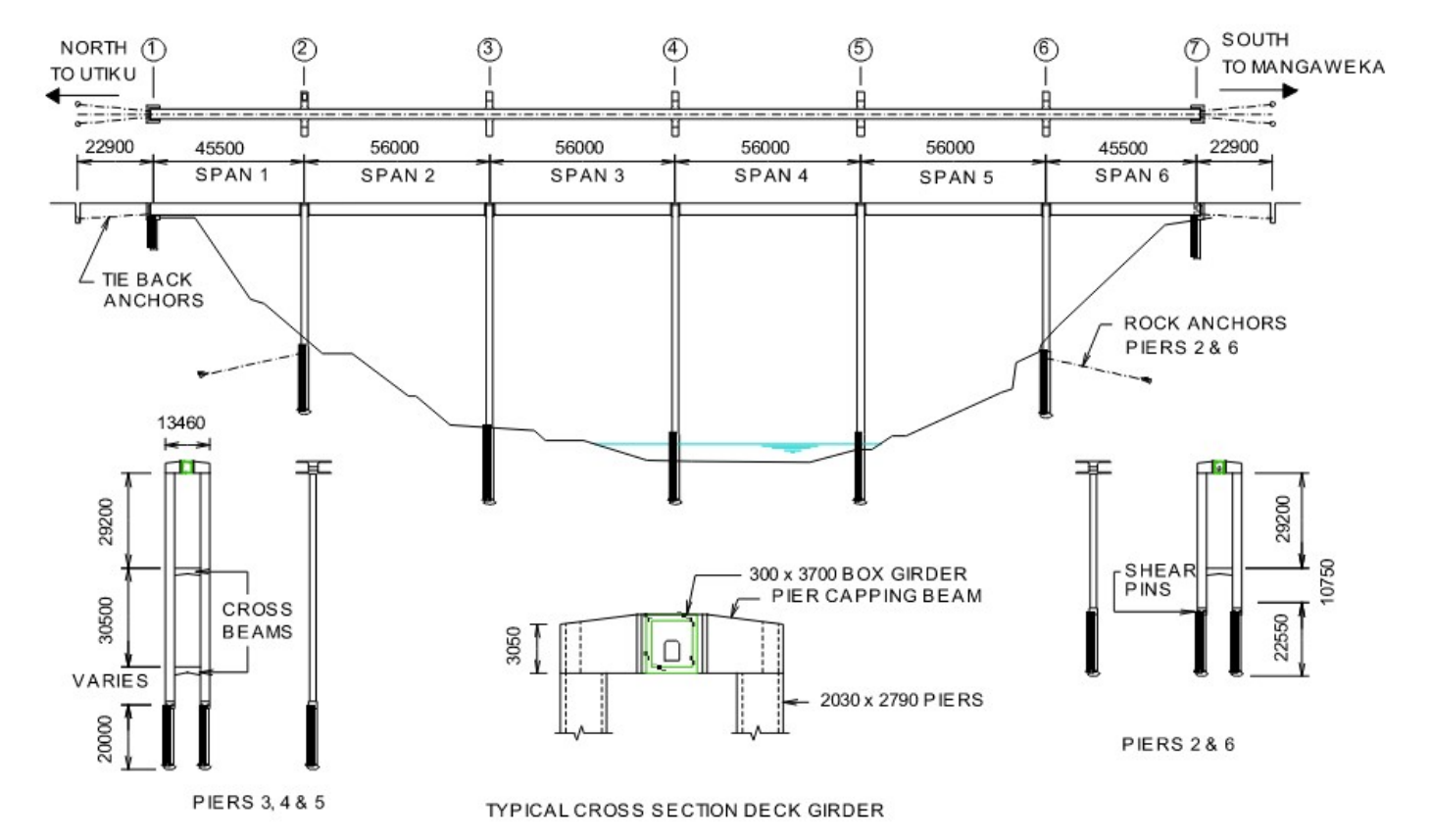}
\par\end{centering}
\caption{\label{fig:South-Rangitikei-viaduct}South Rangitikei Viaduct in New
Zealand (image courtesy of Wikimedia Commons/D B W), bridge and pier
elevations, and cross section of the deck \cite{S_Rangitikei_Const}. }

\end{figure}

Permanent offset (cumulative post-elastic cyclic deformations) in
the response of engineering structures is correlated with increased
repair costs and reduced functionality. Rocking response can reduce
permanent offset. In fact, engineers have long devised rocking mechanisms
in buildings and bridges, despite lack of clear design specifications,
in order to benefit from the reduced design forces when proportioning
tall piers or walls. The motivation is to gain lower ductility demands
in members and connections of the lateral force resisting systems
that in return diminish post-earthquake damage, and to lower the weight
of footings and deep foundation elements \cite{Mar,Beck-1,Panian,Skinner-1}.
These reductions are due to natural period elongation, as shown by
Housner \cite{Housner-2,Housner-3}, Beck and Skinner \cite{Beck-1},
as well as the increased damping during the course of stepping response.
An early example that is operating today is the 315 m long S. Rangitikei
Viaduct with its 76 m tall stepping piers shown in Fig. \ref{fig:South-Rangitikei-viaduct}
\cite{Makris-5}. Energy dissipating devices implanted at the base
in each pier absorb kinetic energy of any induced stepping motion
by plastic deformation of steel torsion beams while providing a safe
stop mechanism to limit liftoff during exceptional loading events
\cite{Beck-1}. The devices were tested for a capacity of 450 kN (101.2
kip) and a range of movement up to 0.08 m (3.15 in or 0.001 drift)
at the Physics and Engineering Laboratory of the former Department
of Scientific and Industrial Research of New Zealand \cite{Beck-1,Kelly-1,Skinner-2},
as well as at the University of California, Berkeley \cite{Kelly-2}.

\subsection{Background}

Most studies to date have focused on dynamic response of rigid bodies,
as discussed in this literature review, to enable predictable practical
limits for tipping. The pervasiveness of investigations of toppling
rigid bodies is likely due to several important applications; for
instance, the free-standing columns in historic structures dating
back to antiquity \cite{Zhong} and protection of museum and art objects
\cite{DeJong-5}. Various loading conditions \cite{Vassiliou-1,Vamvatsikos-2,Trifunac,Makris-1,Psycharis},
modeling assumptions (mass distribution \cite{DeJong-2}, rotational
inertia \cite{Makris-4}, soil behavior and foundation impact \cite{Makris-1,DeJong-4}),
as well as various analysis methods (nonlinear static \cite{Reinhorn-1},
incremental dynamic \cite{Vamvatsikos-1}, displacement-based \cite{Vassiliou-2},
spectrum methods \cite{Makris-3}, and similarity laws \cite{DeJong-3,Rosakis-1})
have been considered. More challenging cases for analysis of rocking
flexible structures supported by rigid or flexible foundations have
only been investigated occasionally (see for instance \cite{DeJong-1,DeJong-2,Beck-1,Meek,Housner-3}). 

The majority of these investigations are numerical in nature with
only a few considering pertinent nonlinear behavior from a dynamical
systems perspective; the notable examples being the work of Bruhn
and Koch \cite{Bruhn}, Hogan \cite{Hogan-3,Hogan-1,Hogan-2}, and
Plaut et al. \cite{Plaut-1}. As a prelude to understanding intriguing
patterns of response of rectangular rigid blocks under different earthquake
ground motion amplitudes and frequencies, Hogan studied the steady-state
response and stability criteria of such objects analytically. A counterintuitive
observation of rocking response under forced vibrations beyond the
point of static stability (toppling) was noted in his studies; it
was also observed in the analyses reported here and will be discussed
later in this paper. Bruhn and Koch \cite{Bruhn} noted heteroclinic
orbits and symmetry breaking bifurcations in response of rocking blocks
under a periodic forcing function. 

Motivated by observations of the response of flexible buildings and
liquid tanks during earthquakes, Psycharis \cite{Psycharis} showed
that foundation uplift and interaction with soil increases the fundamental
period but leaves the second and higher modes of vibration of the
superstructure unaffected. The emergence of vertical oscillations
in response, even when the structure is subjected to purely horizontal
excitation, was also noted. In a study of seven rigid blocks, the
influence of the vertical component of ground motion rocking response
was unclear \cite{Vamvatsikos-2}. Lee and Trifunac \cite{Trifunac}
developed a method for estimation of the rocking component associated
with the incident plane \emph{P} and \emph{SV} waves from the known
translational components of strong ground motion. The rocking ground
motions were also studied by Basu et al. \cite{Constantinou-1} where
they showed that under certain assumptions, and for some structures,
the rotational component of ground motion can influence structural
response. The rocking ground motions may be used in multi-component
excitation analysis of structures; although the general consensus
since their development has been that their effects, at least for
short structures, are minor compared to the translational components
of ground motion. The multi-component excitation analyses require
reliable characterization of rotational component of ground motion
which is often obtained from the translational components of acceleration
time series in a closely-spaced network of stations but such characterization
is difficult due to lack of direct instrumental measurement. Efforts
to bypass this limitation has been made recently \cite{Igel-1}. Similar
to other published work in this area, we only consider the horizontal
component of ground motion in our analyses. 

The assumptions of sufficient friction to prevent sliding and perfectly
plastic impacts have been examined for free-standing blocks subject
to artificial earthquake records by Shenton and Jones \cite{Jones}.
They demonstrated that the assumption of sufficient friction is generally
reasonable, especially for tall blocks, with a required minimum friction
coefficient to sustain pure rocking approximated as 0.75$\nicefrac{L_{F}}{H_{r}}$
(three quarters of the width-to-height ratio of the block) \cite{Jones}. 

Stability of rocking response of rigid blocks by Incremental Dynamic
Analysis (IDA) was studied by Lachanas and Vamvatsikos \cite{Vamvatsikos-1}
where a more frequent occurrence of resurrections and higher variability
in response was noted. IDA resurrections refer to regions of presumed
stability, post unstable dynamic response, as intensity of excitation
is increased. The original definition also extends to cases of reversal
in the rate of a measure of dynamic response (such as peak lateral
displacement) with intensity of excitation (that is, smaller displacements
under larger excitation) \cite{Cornell-1}. It is not clear from the
study of IDA curves what role bifurcations play in the observation
of resurrections or if indeed the resurrections are dynamic bifurcations.
The paper \cite{Vamvatsikos-1} also questions whether the onset of
resurrections should be used in establishing stability in a rocking
rigid block, a question that can be addressed directly in the study
of bifurcation of dynamical systems. Since statistical distributions
of ground motion frequency domain properties (such as period associated
with the maximum spectral acceleration) alter with increasing intensities,
it is unclear how simple scaling of amplitudes of ground motion without
regards to changes in the frequency content might bias the distribution
of IDA curves. The ambiguities and biases can be fundamentally avoided
by a proper choice of an intensity measure, an objective that we will
pursue later.

\section{Quasi-dynamic Response}

Equipped with observations of past rocking structures and noting that
considerable economic saving can be achieved by allowing lifting foundations,
Priestley et al. \cite{Priestley-1,Priestley-2} developed a simple
analytical design procedure to estimate the maximum rocking displacement
demand using equivalent linearization and response spectrum techniques
in an iterative trial-and-error method. It is noteworthy that this
early development alluded to limiting structural damage (Figure 1,
page 147 in \cite{Priestley-1}), a principal in modern seismic resiliency
\cite{Krawinkler-1}, as a compelling argument in favor of rocking
structures compared to other classical means of seismic resistance.
The work by Priestley et al. also appears to be the basis of the recent
developments in the adoption of rocking response into seismic design
provisions \cite{AASHTO-2}. However, an issue of occasional lack
of convergence of the numerical solution of demand displacements in
the iterative procedure was noted by Priestley et al. \cite{Priestley-2};
``In some cases no stable response can be achieved.'' We now describe
the design procedure and present analytical solutions, along with
an analysis of their stability, making the original iterative procedure
unnecessary. 

\subsection{Equivalent Linearization of a Single Degree of Freedom Oscillator\label{sec:Equivalent-Linearization-of}}

\begin{figure}
\begin{centering}
\includegraphics[scale=0.8]{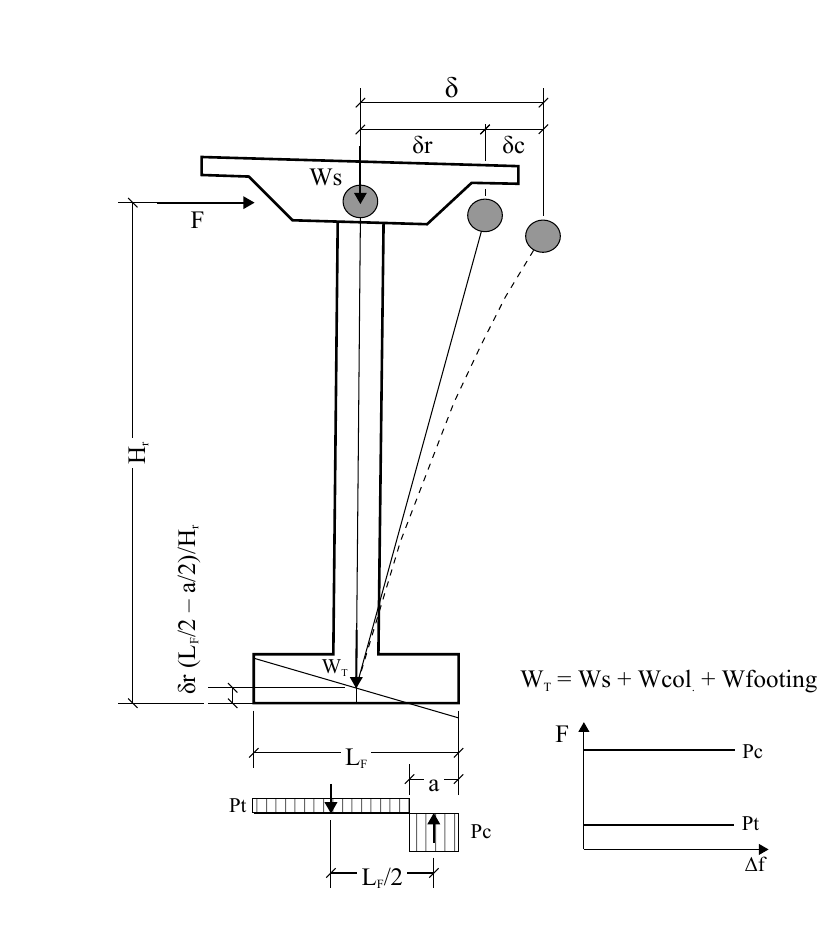}
\par\end{centering}
\caption{\label{fig:Rocking-Pier}Rocking of a single column \cite{Priestley-2}.}
\end{figure}

Priestley's method \cite{AASHTO-2,Priestley-2} is an iterative procedure
for solution of the peak rocking displacement of an equivalent Single
Degree of Freedom (SDOF) oscillator that has three important features:
static equilibrium at its displaced position with the foundation soil
at its limit capacity -see \eqref{eq:1.1}; SDOF oscillator frequency
-see \eqref{eq:1.2}; and pseudo-acceleration design spectrum -see
\eqref{eq:1.3}. With reference to Fig. \ref{fig:Rocking-Pier}:

\begin{equation}
F_{i}=K_{i}\delta_{i}=W_{T}\frac{\left(L_{F}-a\right)}{2H_{r}}-W_{s}\frac{\delta_{i}}{H_{r}}\label{eq:1.1}
\end{equation}
from which $K_{i}$ can be calculated from an initial value of $\delta_{i}$
and substituted into:

\begin{equation}
T_{i}=2\pi\sqrt{\frac{W_{s}+0.5W_{col.}}{g\cdot K_{i}}}\label{eq:1.2}
\end{equation}
to arrive at an updated estimate of $\delta_{i}$ from:

\begin{equation}
\delta_{i+1}=\left(\frac{T_{i}}{2\pi}\right)^{2}\beta\cdot S_{a}\left(T_{i}\right)\label{eq:1.3}
\end{equation}
Here $F_{i}$, $K_{i}$, $\delta_{i}=\delta_{r}+\delta_{c}$, and
$T_{i}$ are the applied force, the lateral bending stiffness, the
total rocking displacement and column deformation, and the period
of vibration at iterate $i$, respectively; $W_{s}$ and $W_{col.}$
are the seismic weight of the superstructure and the weight of the
columns; $W_{T}$ is the total weight of the system; $H_{r}$ is the
height to the centroid of the rocking mass; $L_{F}$ and $B_{F}$
are the length and the width of the footing with $a=W_{T}\left(B_{F}q_{n}\right)^{-1}$,
the width of the rectangular compression stress block at the soil's
capacity under the footing ($P_{c}=q_{n}$); $S_{a}$, $\beta$, and
$g$ are the spectral acceleration of a five percent damped design
spectrum at the site, the spectral acceleration reduction factor due
to damping, and the acceleration due to gravity, respectively. Priestley
\cite{Priestley-2} provides a relationship for $\beta$. Starting
with a chosen initial displacement value $\delta_{0}$, steps \eqref{eq:1.1}
to \eqref{eq:1.3} are recursively evaluated, if the numerical procedure
is stable, until convergence. The converged values of $\delta_{i}$
and $T_{i}$ are denoted $\delta^{*}$ and $T^{*}$ in what follows. 
\begin{assumption*}
\textbf{\textup{Equivalent Linearization:}} The rocking oscillation
is inherently nonlinear due to the period of vibration dependence
on displacement and the sudden change of restoring moment at full-base
contact \cite{Beck-1,Housner-2,Pellegrino}. The equivalent linearization
method is an approximation to the solution of the equation of motion
\cite{Caughey-1}. SDOF response is assumed with no interaction from
superstructure rotational stiffness. The soil-footing interface is
assumed rigid perfectly-plastic. Sections \ref{subsec:Example-2}
and \ref{subsec:Example-4} exhibit the influence of equivalent linearization
on the predicted response of a typical bridge bent.
\end{assumption*}

\subsection{Analytical Design Demand Solutions and Their Stability for Linearized
Response}

\begin{figure}
\begin{centering}
\includegraphics[scale=0.5]{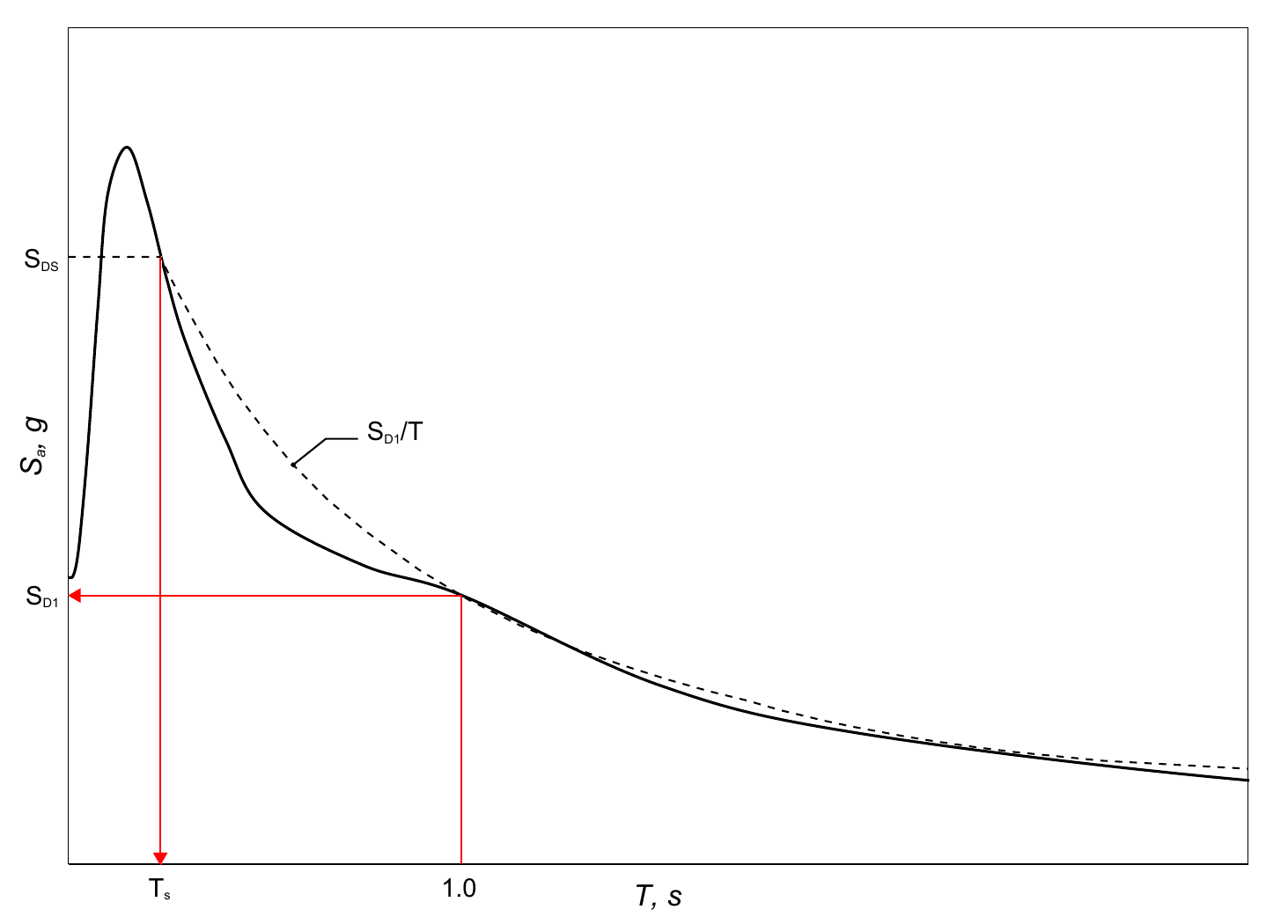}
\par\end{centering}
\caption{\label{fig:A-two-period-seismic}A multi-period design spectrum (solid
curve) and its simplification (dashed curve). }

\end{figure}

Equations \eqref{eq:1.1} to \eqref{eq:1.3} can be combined to express
the stepping response of a linearized SDOF oscillator in the form
of a nonlinear map. Nonlinear maps, or difference equations, are a
class of dynamical systems in which time is discrete and they are
often used in the analysis of differential equations \cite{Strogatz-1}.
In what follows, a traditional ``two-period'' design response spectrum
\cite{AASHTO-2,ASCE-1,ASCE-2} is assumed for consistency with seismic
hazard and ground motion characterization in the procedures of Displacement-based
Seismic Design \cite{AASHTO-1,Priestley-3}. An example of a two-period
acceleration design response spectrum is shown in Fig. \ref{fig:A-two-period-seismic}
where $T_{s}$, the characteristic period of ground motion at the
intersection of constant acceleration and constant velocity segments
of the design spectrum \cite{Clough-2,Fajfar-1}, is given by $\nicefrac{S_{D1}}{T_{s}}=S_{a}\left(T_{s}\right)=S_{DS}$,
so

\[
T_{s}\triangleq\frac{S_{D1}}{S_{DS}}
\]
and $S_{D1}$ and $S_{DS}$ are the design spectral response shape
parameters at 1.0 s and ``short periods,'' respectively. Regardless
of the shape of the design spectrum (e.g., multi-period design response
spectrum \cite{ASCE-2} or site-specific spectra), the analysis of
nonlinear maps to study convergence for design displacements is relevant;
the approach in §\ref{subsec:The-Iterated-Maps} may be taken as illustration
of such treatment.

\subsubsection{The Iterated Maps of Design Displacement\label{subsec:The-Iterated-Maps}}
\begin{condition*}
\textbf{(a) ``Short-period'' Structure,} $\mathbf{T_{i}\leq T_{s},\,S_{a}=S_{DS}}$:
Rearranging \prettyref{eq:1.1} to \prettyref{eq:1.3} gives:
\end{condition*}
\begin{equation}
\delta_{i+1}=\lambda_{1}\left(\frac{\delta_{i}}{1-w\delta_{i}}\right)\triangleq f_{1}\left(\delta_{i}\right)\label{eq:2.1}
\end{equation}
where

\begin{equation}
\lambda_{1}=\frac{\beta S_{DS}}{g}\cdot\frac{W_{s}+0.5W_{col.}}{W_{T}\frac{\left(L_{F}-a\right)}{2H_{r}}}\simeq\frac{2\beta S_{DS}H_{r}}{g\left(L_{F}-a\right)}\label{eq:2.2}
\end{equation}
and

\begin{equation}
w=\frac{2W_{s}}{W_{T}\left(L_{F}-a\right)}\label{eq:2.3}
\end{equation}
$\lambda_{1}\in\mathbb{R^{\mathrm{+}}}$ is dimensionless, typically
in $\left[10^{-5},10^{-1}\right]$, and $w\in\mathbb{\mathbb{R^{\mathrm{+}}}}$
has unit $\left[L^{-1}\right]$. Note that $\delta_{i}<w^{-1}\simeq L_{F}-a$
for a physically viable solution. 

\begin{condition*}
\textbf{(b) ``Long-period'' Structure,} $\mathbf{T_{i}>T_{s},\,S_{a}=S_{D1}\cdot T_{i}^{-1}}$:
Rearranging \prettyref{eq:1.1} to \prettyref{eq:1.3} gives:
\end{condition*}
\begin{equation}
\delta_{i+1}=\lambda_{2}\sqrt{\frac{\delta_{i}}{1-w\delta_{i}}}\triangleq f_{2}\left(\delta_{i}\right)\label{eq:2.4}
\end{equation}
where

\begin{equation}
\lambda_{2}=\frac{\beta S_{D1}}{2\pi\sqrt{g}}\left[\frac{W_{s}+0.5W_{col.}}{W_{T}}\cdot\frac{2H_{r}}{\left(L_{F}-a\right)}\right]^{\frac{1}{2}}\simeq\frac{\sqrt{2}}{2}\left(\frac{\beta S_{D1}}{\pi\sqrt{g}}\right)\sqrt{\frac{H_{r}}{L_{F}-a}}\label{eq:2.5}
\end{equation}
$\lambda_{2}\in\mathbb{R^{\mathrm{+}}}$ has unit $\left[L\right]^{\nicefrac{1}{2}}$,
typically in $\left[10^{-4},10^{0}\right]$. Note that $\delta_{i}<w^{-1}$
for a physically viable solution, and the scalars $\lambda_{1}$ and
$\lambda_{2}$ characterize the severity of excitation with:

\begin{equation}
\frac{\lambda_{1}}{\lambda_{2}}=\frac{2\sqrt{2}\pi\cdot S_{DS}\cdot\sqrt{H_{r}}}{\sqrt{g\left(L_{F}-a\right)}\cdot S_{D1}}=2\pi\left(\frac{S_{DS}}{S_{D1}}\right)\sqrt{\frac{2H_{r}}{g\left(L_{F}-a\right)}}\label{eq:2.7}
\end{equation}

\begin{sidewaysfigure}
\begin{centering}
\includegraphics[viewport=0cm 0cm 20cm 30cm,width=0.85\paperwidth]{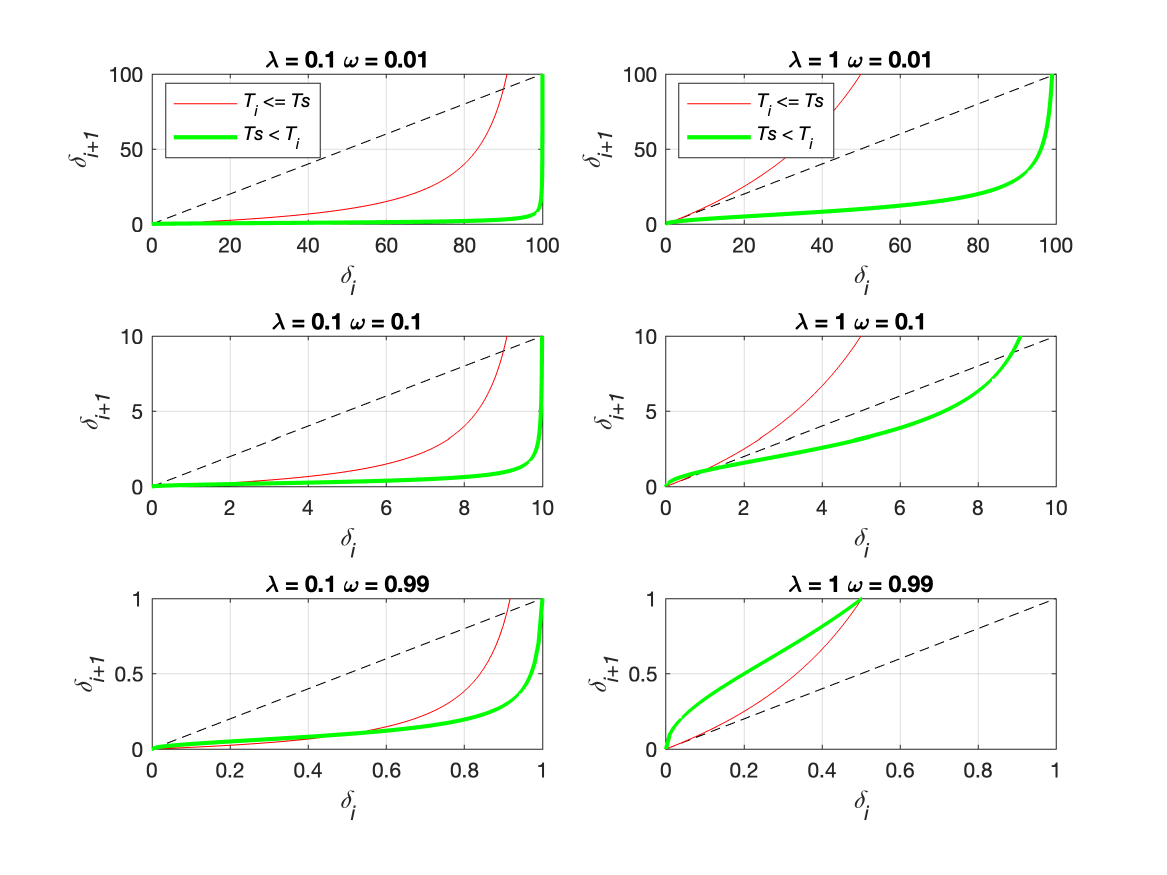}
\par\end{centering}
\caption{\label{fig:The-stepping-maps}The stepping maps for different values
of $\lambda$ and $w$.}
\end{sidewaysfigure}

The iterated maps of \eqref{eq:2.1} and \eqref{eq:2.4} are illustrated
for different values of $\lambda_{1}=\lambda_{2}=\lambda$ and $w$
in Fig. \ref{fig:The-stepping-maps}. Note that the shape of the long-period
map ($T_{s}<T$) resembles that of $\cosh^{-1}$ function; as in the
variation of the period of vibration of rocking blocks with the inverse
of rocking amplitude shown in \cite{Housner-2}.

\subsubsection{Fixed Points of Displacement and Their Linear Stability\label{subsec:Fixed-Points-of}}

\LyXZeroWidthSpace{}

\textbf{(a) \textquotedblleft Short-period\textquotedblright{} Structure:}
For $T_{i}\leq T_{s}$, from \prettyref{eq:2.1}, the fixed point
of $\delta$, called $\delta^{*}$ is:

\begin{equation}
\delta^{*}=f_{1}\left(\delta^{*}\right)=\lambda_{1}\left(\frac{\delta^{*}}{1-w\delta^{*}}\right)\label{eq:2.8}
\end{equation}
or

\begin{equation}
\begin{split}\delta_{1}^{*}=0,\: & \delta_{2}^{*}=\frac{1-\lambda_{1}}{w}\end{split}
\label{eq:2.9}
\end{equation}
$\delta_{1}^{*}$ is a trivial solution at undisturbed equilibrium,
and

\begin{equation}
\delta_{2}^{*}\simeq L_{F}-a-\frac{2\beta S_{DS}H_{r}}{g}<\left(L_{F}-a\right)\label{eq:2.10}
\end{equation}
assuming $2W_{s}\approx W_{T}$. The iterated map $f_{1}$ is a $C^{1}$
smooth function from $\mathbb{R^{\mathrm{+}}}$ to itself for $\delta^{*}<w^{-1}$.
The multiplier for linear stability is:

\begin{equation}
f_{1}'\left(\delta^{*}\right)=\frac{\lambda_{1}}{\left(1-w\delta^{*}\right)^{2}}\label{eq:2.11}
\end{equation}
$\delta_{1}^{*}$ is linearly stable if $\left|f_{1}'\left(0\right)\right|=\lambda_{1}<1$.
The other fixed point, $\delta_{2}^{*}$, is unstable since $\left|f_{1}'\left(\frac{1-\lambda_{1}}{w}\right)\right|=\lambda_{1}^{-1}>1$
for a physically viable $\delta_{2}^{*}$ based on \prettyref{eq:2.9},
that is, the short period structure always converges to the zero displacement
equilibrium as long as:

\begin{equation}
\beta S_{DS}<\frac{W_{T}}{\left(W_{s}+0.5W_{col.}\right)}\cdot\frac{\left(L_{F}-a\right)}{2H_{r}}\cdot g\label{eq:2.12}
\end{equation}

\begin{note*}
$\delta_{2}^{*}$ is an exact solution of \eqref{eq:1.1} - \eqref{eq:1.3}
but the iterative procedure presumably gives $\delta_{1}^{*}=0$ if
$\delta_{0}\leq\delta_{2}^{*}$ and moves away from $\delta_{2}^{*}$
if $\delta_{0}>\delta_{2}^{*}$. Thus, the iterative procedure does
not work for case (a). 
\end{note*}
\textcompwordmark\textbf{(b) ``Long-period'' Structure:} For $T_{s}<T_{i}$,
from \prettyref{eq:2.4}, the fixed point of $\delta$, $\delta^{*}$
is:

\begin{equation}
\delta^{*}=f_{2}\left(\delta^{*}\right)=\lambda_{2}\sqrt{\frac{\delta^{*}}{1-w\delta^{*}}}\label{eq:2.13}
\end{equation}
or

\begin{equation}
\begin{split}\delta_{1}^{*}=0,\: & \delta_{3,4}^{*}=\frac{1\mp\sqrt{1-4w\lambda_{2}^{2}}}{2w}\end{split}
\label{eq:2.14}
\end{equation}
$\delta_{1}^{*}$ is a trivial solution at undisturbed equilibrium
and $f_{2}$ is a $C^{1}$ smooth function from $\mathbb{R}^{+}$
to itself for $\delta^{*}<w^{-1}$. The multiplier for linear stability
is: 

\begin{equation}
f_{2}'\left(\delta^{*}\right)=\left(\frac{\lambda_{2}}{2}\right)\left(\frac{\delta^{*}}{1-w\delta^{*}}\right)^{-\frac{1}{2}}\left(1-w\delta^{*}\right)^{-2}\label{eq:2.15}
\end{equation}

Linear stability of $\delta_{1}^{*}$ is undefined; however, for $\delta_{3,4}^{*}$:

\begin{equation}
f_{2}'\left(\delta_{3,4}^{*}\right)=\pm\frac{\sqrt{1-2w\lambda_{2}^{2}\pm\sqrt{1-4w\lambda_{2}^{2}}}}{2\sqrt{2}w\lambda_{2}^{2}}\label{eq:2.16}
\end{equation}
for:

\begin{equation}
\delta_{3}^{*}=\frac{1-\sqrt{1-4w\lambda_{2}^{2}}}{2w}\label{eq:2.19a}
\end{equation}
its existence and linear stability are guaranteed for $0<w\lambda_{2}^{2}<\frac{1}{4}$
, therefore, this condition for $\delta_{3}^{*}$ can be translated
to an upper limit for intensity of shaking, while $\delta_{4}^{*}=\left(1+\sqrt{1-4w\lambda_{2}^{2}}\right)\left(2w\right)^{-1}$
is unstable because $\left|f_{2}'\left(\delta_{4}^{*}\right)\right|>1$: 

\begin{equation}
\beta S_{D1}<\frac{\pi}{2}\sqrt{\frac{g}{H_{r}}}\left(L_{F}-a\right)\label{eq:2.17}
\end{equation}

Conditions \prettyref{eq:2.12} and \eqref{eq:2.17} refer to stability
of solution in converging to a fixed point of displacement and they
should not be confused with mechanical stability of the structure
under consideration. 
\begin{rem}
The analysis presented here implies that the simplified design procedure
does not converge to any solution in Case (a) and only to one of the
solutions in Case (b). The simplified design procedure, thus, has
limited utility as a general method of estimating demand displacements
in the analysis of flexible stepping frames. While the simplified
procedure may be useful in the early stages of design, the nonlinear
dynamic (time-stepping) analyses shown in §\ref{sec:The-Equations-of}
are recommended for such applications. It is worth noting also that
the fixed points of displacement obtained in §\ref{subsec:Fixed-Points-of}
do not reflect the dynamic response under temporal nonstationarities
of a time-varying forcing function (e.g., moving resonance as described
in \cite{Beck-2,Wen-1} and §\ref{Moving-resonance-}) because the
spectral ordinates in Fig. \ref{fig:A-two-period-seismic} are obtained
from response of a linear oscillator \cite{Hjelmstad-2}. The spectral
ordinate is the peak response obtained from linear dynamic analysis
of an equivalent oscillator with calibrated period of vibration and
it is only an approximation to the peak response of the nonlinear
oscillator. The variation of natural period of vibration during the
iterative process of §\ref{sec:Equivalent-Linearization-of} is a
consequence of the numerical solution procedure rather than the inherent
nonlinear behavior of the ``inverted'' oscillator in which the period
of vibration grows with the amplitude of displacement response. In
fact, Fig. \ref{fig:Convergence-trajectories-from} from Example 2
in §\ref{subsec:Example-2} shows that the period of vibration may
lengthen or shorten during the iterative process of finding a fixed
point, depending on the initial value used to start the iterations. 
\end{rem}

\LyXZeroWidthSpace{}
\begin{rem}
Seismic design codes are moving towards multi-period design spectrum
\cite{ASCE-2} because the simpler piecewise spectral shape that has
been used in the past 40 years \cite{Newmark-1}, parameterized by
only a few points, appears to underestimate spectral demands on soft
soil sites when ground motion hazard is dominated by large magnitude
events. With such spectra (shown in Fig. \ref{fig:A-two-period-seismic}
along with its simpler piecewise approximation), the conditions of
§\ref{subsec:Fixed-Points-of} would apply in principle with some
adjustment. 
\end{rem}

\begin{figure}
\begin{centering}
\includegraphics[scale=0.7]{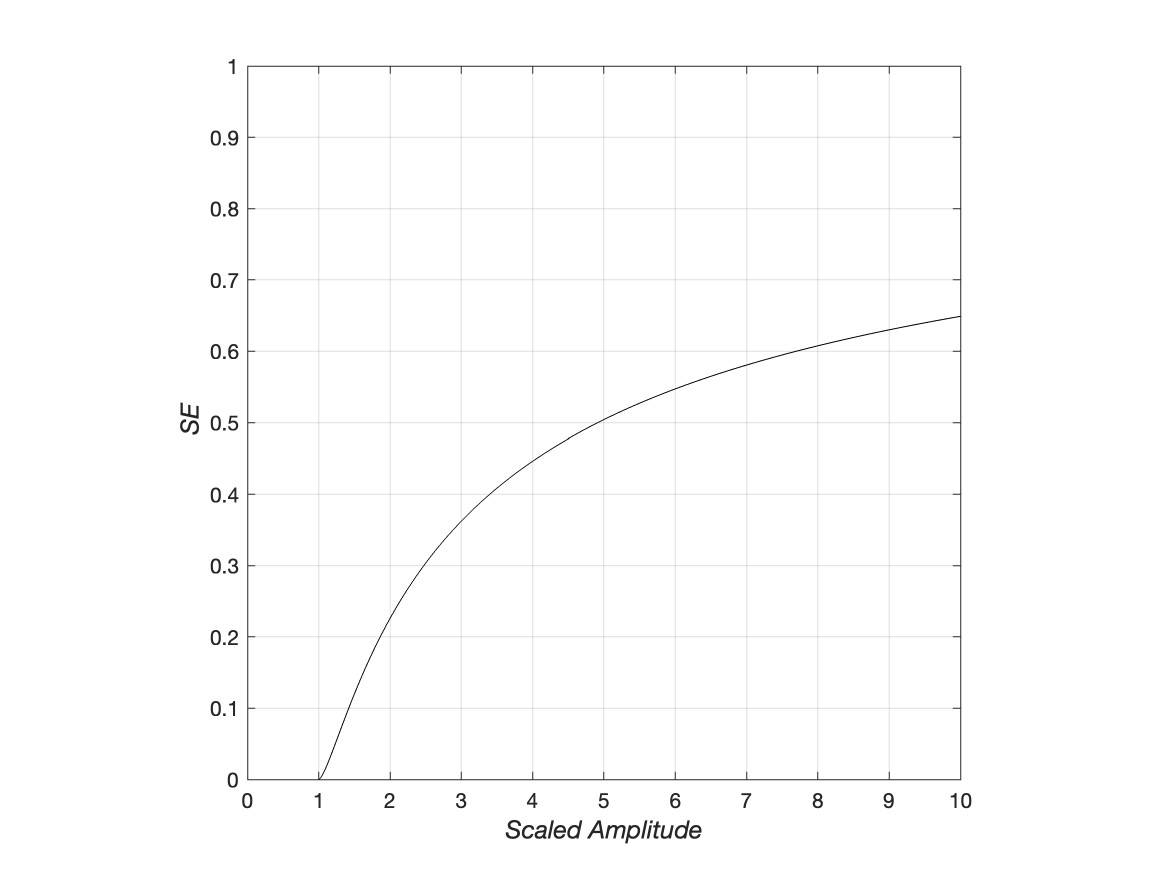}
\par\end{centering}
\caption{\label{fig:Variation-of-SE}Variation of \emph{SE} with the amplitude
of vibration (scaled to lift-off initiation amplitude). }

\end{figure}

The condition of instability for $\delta_{2}^{*}$ stated above is
necessary but not sufficient. It is notable from the iterated maps
shown in Fig. \ref{fig:The-stepping-maps}, and from consideration
of the functional form of \eqref{eq:2.8}, that growing displacements
while iterating may induce a transition from the unstable region of
$T_{i}\leq T_{S}$ to $T_{S}<T_{i}$ in \eqref{eq:1.3}, violating
the period range. From \eqref{eq:1.1} and \eqref{eq:1.2} the growth
ratio of natural period over one iteration is:

\begin{equation}
\frac{T_{i+1}}{T_{i}}=\zeta_{i}\cdot\sqrt{\frac{1-w\delta_{i}}{1-w\delta_{i+1}}}\label{eq:2.18}
\end{equation}
where 

\begin{equation}
\zeta_{i}=\sqrt{\frac{\delta_{i+1}}{\delta_{i}}}\label{eq:2.19}
\end{equation}
and $w$ is defined in \eqref{eq:2.3}. Note for $w\ll1$ we have:

\begin{equation}
T_{i+1}\approx\zeta_{i}\cdot T_{i}\label{eq:2.20}
\end{equation}
\emph{Stepping Effectiveness (SE)} may be defined, similar to isolation
effectiveness in \cite{Clough-2}, as a measure of reduction in design
forces. If $F_{0}$ and $F_{n}$ are the initial and the converged
design force, respectively, considering \eqref{eq:1.1}, \eqref{eq:1.3},
\eqref{eq:2.18}, and \eqref{eq:2.19}, we have:

\begin{multline}
SE\triangleq1-\frac{\underset{i\rightarrow\infty}{\lim}\left(F_{i}\right)}{F_{0}}\cong1-\frac{F_{n}}{F_{0}}=1-\left(\frac{1-w\delta_{1}}{1-w\delta_{0}}\right)\left(\frac{1-w\delta_{2}}{1-w\delta_{1}}\right)\cdots\left(\frac{1-w\delta_{n}}{1-w\delta_{n-1}}\right)=\\
1-\stackrel[i=0]{n}{\prod}\left[\left(\frac{T_{i}}{T_{i+1}}\right)\zeta_{i}\right]^{2}=1-\left(\frac{T_{0}}{T_{n}}\right)^{2}\stackrel[i=0]{n}{\prod}\zeta_{i}^{2}=1-\left(\frac{T_{0}}{T_{n}}\right)^{2}\left(\frac{T_{n}}{T_{0}}\right)=1-\frac{T_{0}}{T_{n}}\label{eq:2.21}
\end{multline}
that is, \emph{SE} depends on the elongation of the natural period
of vibration of the oscillator. In terms of the amplitudes of vibration,
the effectiveness of stepping response improves with increase in the
amplitudes of vibration \cite{Beck-1}, as shown in Fig. \ref{fig:Variation-of-SE}.
The challenge in practical situations is maintaining stability while
increasing the effectiveness of stepping response. 

\LyXZeroWidthSpace{}

\textcompwordmark{}

\subsubsection{Example 1:\label{subsec:Example-1} A Reinforced Concrete Single-column
Pier}

\LyXZeroWidthSpace{}

We consider a bridge bent shown in Fig. \ref{fig:Rocking-Pier} (Example
6.6, page 525 in \cite{Priestley-2}) with:
\begin{itemize}
\item The weight of the superstructure $W_{s}=8000$ kN,
\item The weight of the column $W_{col.}=424$ kN,
\item The gross weight of the pier $W_{T}=9600$ kN, 
\item $L_{F}=B_{F}=7$ m, 
\item $q_{n}=1$ MPa, 
\item $H_{r}=27$ m, and
\item The initial, ``small-displacement '' natural period, $T_{1}=4.34$
s,
\end{itemize}
at a site characterized by:
\begin{itemize}
\item $\beta S_{D1}=5.2492$ $\mathrm{m\cdot s^{-1}}$, and
\item $T_{s}=0.62$ s.
\end{itemize}
Three iterations in \cite{Priestley-2} arrive at $\delta=0.722$
m after which convergence is assumed. 
\begin{sol*}
Given that $T_{s}<T_{1}$, we first check the stability of solution
in \eqref{eq:2.16}. With $a=\frac{W_{T}}{B_{F}q_{n}}=\frac{9600\times10^{3}}{7\left(10^{6}\right)}=1.3714$
m, we calculate

\[
w_{2}=\frac{2W_{s}}{W_{T}\left(L_{F}-a\right)}=\frac{2\left(8000\times10^{3}\right)}{9600\times10^{3}\left(7-1.3714\right)}=0.0296\,\mathrm{m^{-1}}
\]
and

\begin{multline*}
\lambda_{2}=\frac{\beta S_{D1}}{2\pi\sqrt{g}}\left[\frac{W_{s}+0.5W_{col.}}{W_{T}}\cdot\frac{2H_{r}}{\left(L_{F}-a\right)}\right]^{\frac{1}{2}}=\\
\frac{5.2492}{2\left(3.1416\right)\sqrt{9.81}}\left[\frac{8000+\frac{1}{2}\left(212\right)}{9600}\cdot\frac{\left(2\right)\left(27\right)}{\left(7-1.3714\right)}\right]^{\frac{1}{2}}=0.7592\,\mathrm{m^{\frac{1}{2}}}
\end{multline*}
Since $0<\left(w_{2}\lambda_{2}^{2}=0.0171\right)<\frac{1}{4}$, from
\eqref{eq:2.14}, the fixed point $\delta_{3}^{*}=\frac{1-\sqrt{1-4\left(0.0296\right)\left(0.7592\right)^{2}}}{\left(2\right)\left(0.0296\right)}=0.5865$m
is a numerically stable solution of the iterative procedure. 
\end{sol*}

\subsubsection{Example 2: \label{subsec:Example-2} A Two-column Steel Frame}

\LyXZeroWidthSpace{}

\begin{figure}
\begin{centering}
\includegraphics[scale=0.65]{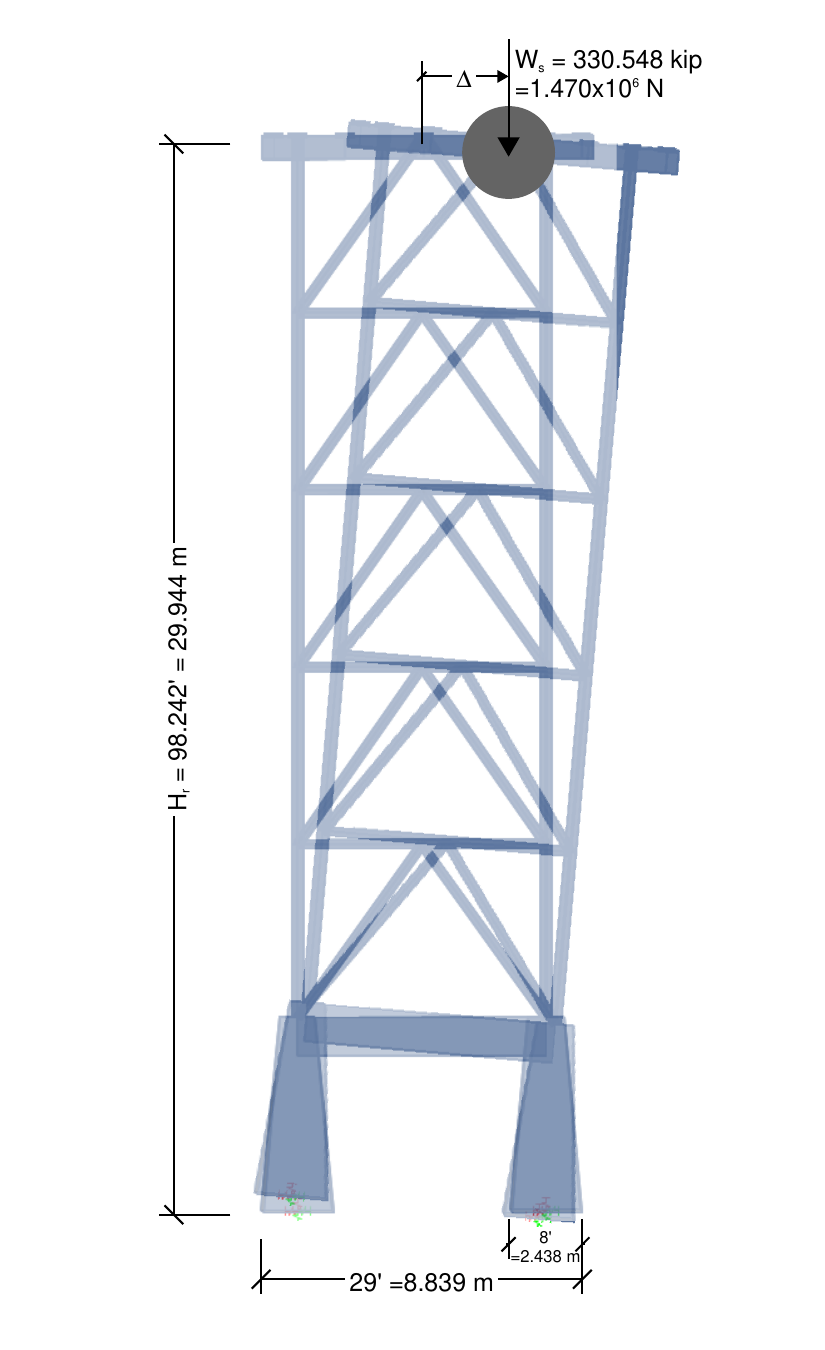}
\par\end{centering}
\caption{\label{fig:Stepping-frame-of}Stepping frame of Example 2.}
\end{figure}

Consider the frame shown in Fig. \ref{fig:Stepping-frame-of} \cite{Alimoradi-1}.
It had been observed in an earlier study that designing the frame
for the required seismic forces had made the size and the cost of
the footings prohibitive. We calculate the design displacement $\delta_{3}^{*}$
of the frame following the procedures of \cite{AASHTO-2,Priestley-2},
described under §\ref{sec:Equivalent-Linearization-of}, and then
the closed-form solution obtained in §\ref{subsec:The-Iterated-Maps}
with:
\begin{itemize}
\item The weight of the superstructure $W_{s}=331$ kip $=1.472\times10^{6}$
N, 
\item The gross weight of the bent $W_{T}=631$ kip $=2.807\times10^{6}$
N,
\item $L_{F}=29$ ft $=8.839$ m, 
\item $a=3.824$ ft $=1.166$ m, and
\item $H_{r}=98.242$ ft $=29.944$ m,
\end{itemize}
at a site characterized by:
\begin{itemize}
\item $S_{DS}=1.038$ g $=10.180$ $\mathrm{m\cdot s^{-2}}$, 
\item $S_{D1}=0.456$ g $=4.472$ $\mathrm{m\cdot s^{-1}}$, and
\item $T_{s}=\frac{S_{D1}}{S_{DS}}=0.439$ $\mathrm{s}$.
\end{itemize}
Assume $\beta=0.8$ for 20 percent reduction in spectral acceleration
due to damping. 

\begin{figure}
\begin{centering}
\includegraphics[scale=0.7]{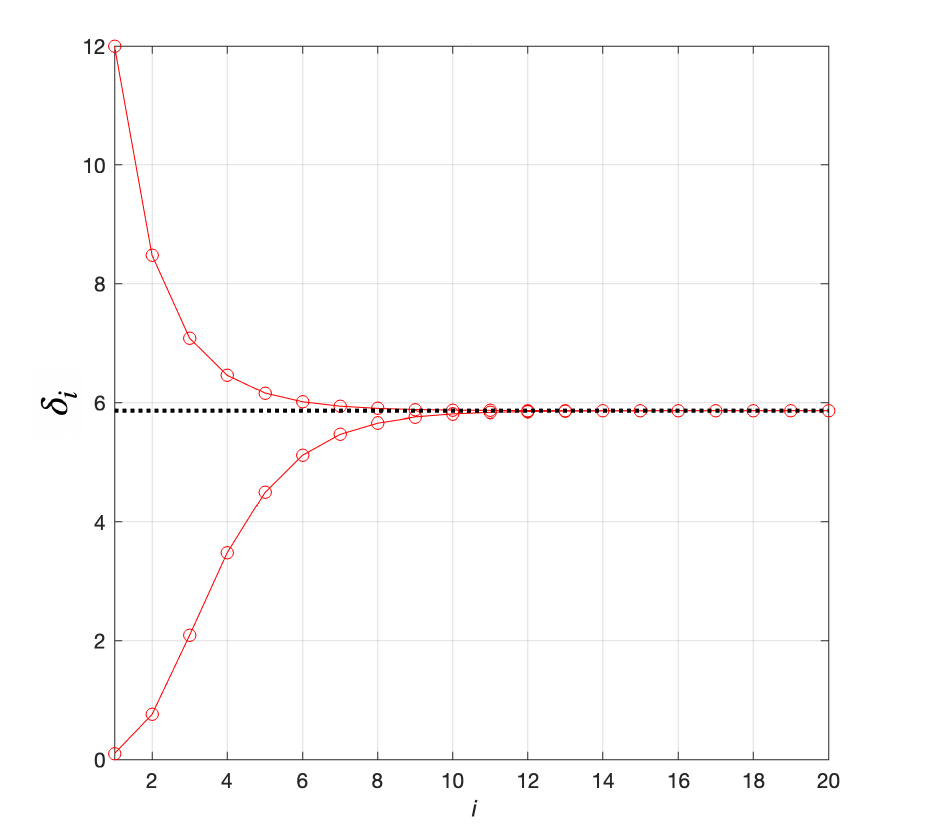}
\par\end{centering}
\caption{\label{fig:Convergence-trajectories-from}Convergence trajectories
from two initial values in Example 2.}
\end{figure}

\begin{sol*}
Table \ref{tab:Comparison-of-AASHTO} and Fig. \ref{fig:Convergence-trajectories-from}
present the results of the iterative procedure for two arbitrary initial
values and the closed-form solution for $\delta_{3}^{*}$ from \eqref{eq:2.14}.
The dependence of the iterative procedure on initial values is avoided
in the procedures of §\ref{subsec:Fixed-Points-of}. For this problem,
the parameters are:
\end{sol*}
\begin{itemize}
\item $w_{2}=0.042\,\mathrm{ft^{-1}=0.138\,m^{-1}}$ and
\item $\lambda_{2}=0.692\:\mathrm{ft^{\frac{1}{2}}=0.382\,m^{\frac{1}{2}}}$.
\end{itemize}
The equivalent linear period corresponding to $\delta_{3}^{*}$ is
$T^{*}=1.643\mathrm{s}>T_{s}$, giving consistency with the use of
the ``long-period'' structure expression \eqref{eq:2.14}.

\begin{table}
\caption{\label{tab:Comparison-of-AASHTO}Comparison of AASHTO Appendix A procedure
with $w_{2}=0.0416774$ and $\lambda_{2}=0.691857$. Values in inches
(1 in = 0.0254 m.) }

\begin{centering}
\begin{tabular}{|c|c|c|c|}
\hline 
\emph{i} & $\delta_{i}$ from AASHTO (Case I) & $\delta_{i}$ from AASHTO (Case II) & $\delta_{3}^{*}$ from \eqref{eq:2.14}\tabularnewline
\hline 
\hline 
1 & 1.0000000e-01 & 1.2000000e+01 & 5.8622044e+00\tabularnewline
\hline 
2 & 7.5794068e-01 & 8.4799348e+00 & \tabularnewline
\hline 
3 & 2.0890540e+00 & 7.0834703e+00 & \tabularnewline
\hline 
4 & 3.4762864e+00 & 6.4578903e+00 & \tabularnewline
\hline 
5 & 4.4952588e+00 & 6.1592793e+00 & \tabularnewline
\hline 
6 & 5.1209834e+00 & 6.0120048e+00 & \tabularnewline
\hline 
7 & 5.4718291e+00 & 5.9381421e+00 & \tabularnewline
\hline 
8 & 5.6596764e+00 & 5.9007789e+00 & \tabularnewline
\hline 
9 & 5.7579193e+00 & 5.8817960e+00 & \tabularnewline
\hline 
10 & 5.8086891e+00 & 5.8721299e+00 & \tabularnewline
\hline 
11 & 5.8347664e+00 & 5.8672023e+00 & \tabularnewline
\hline 
12 & 5.8481192e+00 & 5.8646888e+00 & \tabularnewline
\hline 
13 & 5.8549456e+00 & 5.8634064e+00 & \tabularnewline
\hline 
14 & 5.8584326e+00 & 5.8627519e+00 & \tabularnewline
\hline 
15 & 5.8602132e+00 & 5.8624179e+00 & \tabularnewline
\hline 
16 & 5.8611221e+00 & 5.8622475e+00 & \tabularnewline
\hline 
17 & 5.8615861e+00 & 5.8621605e+00 & \tabularnewline
\hline 
18 & 5.8618229e+00 & 5.8621161e+00 & \tabularnewline
\hline 
19 & 5.8619438e+00 & 5.8620934e+00 & \tabularnewline
\hline 
20 & 5.8620055e+00 & 5.8620819e+00 & \tabularnewline
\hline 
\end{tabular}
\par\end{centering}
\end{table}

\section{Dynamic Response}

\subsection{\label{sec:The-Equations-of}The Equations of Motion for Nonlinear
Stepping Response}

Attempts at understanding the dynamics of flexible lifting frames
were made at University of Tokyo with experiments as early as 1960
\cite{Muto-1}. In the 1970s, the Department of Scientific and Industrial
Research of New Zealand developed equations of motion and analyzed
the seismic response of a flexible stepping A-frame pier that vibrated
between unstepped and stepped phases for sufficiently large ground
shaking \cite{Beck-1}. A review of the basics of the procedure follows
to facilitate extension to the equations of motion of a flexible portal
frame.

\subsubsection{Stepping A-Frame Dynamics}

\begin{figure}
\begin{centering}
\includegraphics[scale=0.8]{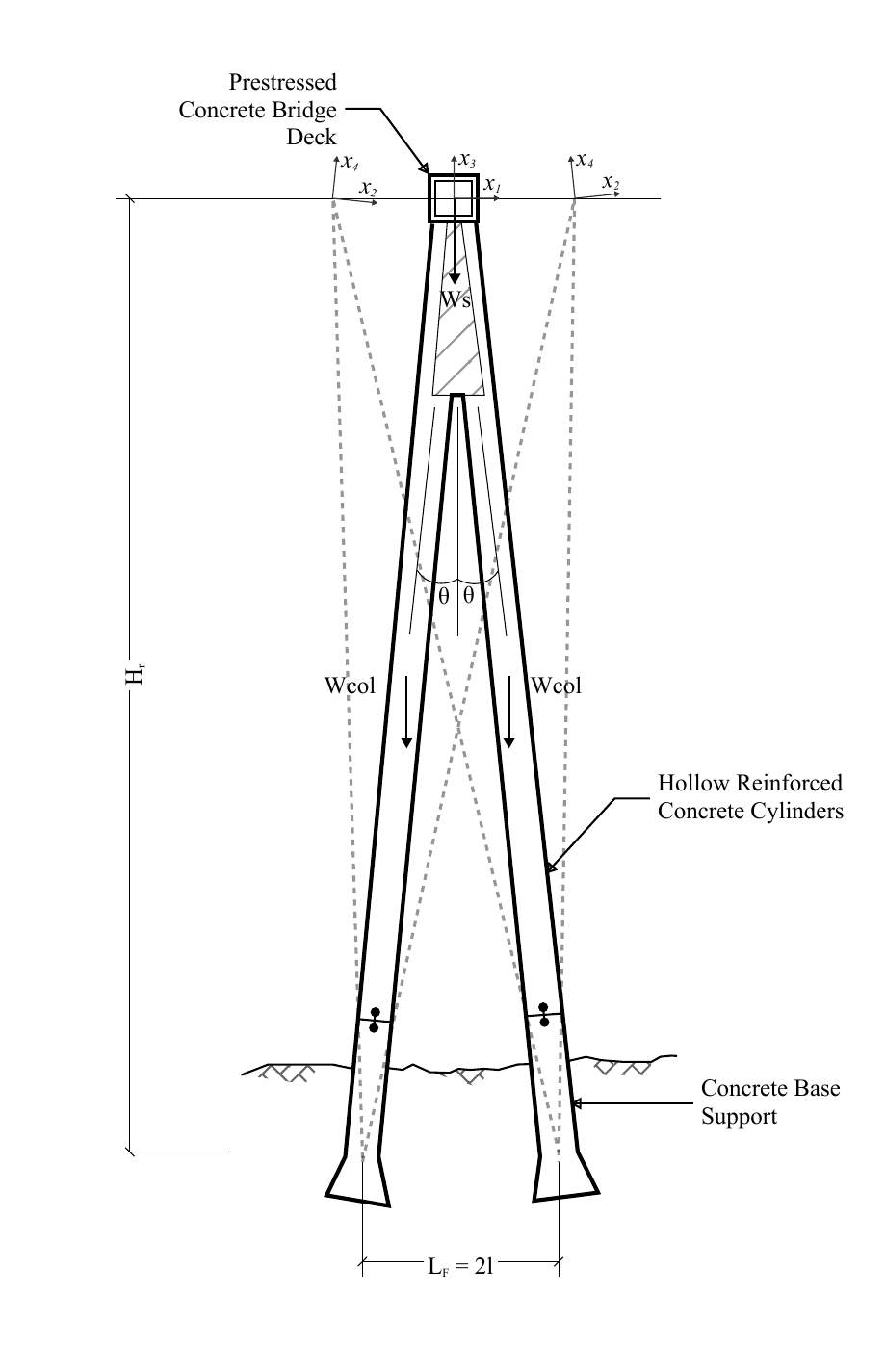}
\par\end{centering}
\caption{\label{fig:Stepping-pier-.}Stepping ``A'' frame (after \cite{Beck-1}).}
\end{figure}

Using d'Alembert's principle while noting that bending and axial oscillations
of the columns of uniform mass contribute insignificantly to the overall
motion of an A-frame (Fig. \ref{fig:Stepping-pier-.}) and using a
model with only two degrees of freedom at the apex, the equations
of motion for a stepping A-frame subject to ground acceleration $\ddot{x}_{g}\left(t\right)$,
are \cite{Beck-1}:
\begin{equation}
\left\{ \begin{array}{c}
\ddot{x}_{1}+\pi\xi_{1}\omega_{1}^{2}\,\mathrm{sign}\left(\dot{x}_{1}\right)\left|x_{1}\right|+\omega_{1}^{2}x_{1}=-P_{1}\ddot{x}_{g}\\
\ddot{x}_{3}+\omega_{3}^{2}x_{3}=0.
\end{array}\right.\begin{array}{c}
\quad\quad\mathrm{\qquad\qquad if}\end{array}\left|x_{1}\right|\leq x_{c}\label{eq:4.1}
\end{equation}

\begin{equation}
\left\{ \begin{array}{c}
\ddot{x}_{2}-\omega_{2}^{2}\sec\left(\theta\right)x_{1}+\gamma_{1}\sec\left(\theta\right)\mathrm{sign}\left(x_{1}\right)=-P_{2}\sec\left(\theta\right)\ddot{x}_{g}\\
\ddot{x}_{4}+\omega_{4}^{2}x_{4}-\gamma_{2}=P_{4}\,\mathrm{sign}\left(x_{1}\right)\ddot{x}_{g}
\end{array}\right.\begin{array}{c}
\mathrm{if}\end{array}\left|x_{1}\right|>x_{c}\label{eq:4.2}
\end{equation}

At the changeover from one phase of motion to the other, the condition
of continuity of velocity (or lack thereof) in \eqref{eq:4.3} provides
the initial conditions for the next phase:

\begin{equation}
\dot{x}_{2}\left(t_{I}^{+}\right)=C_{r}\,\dot{x}_{1}\left(t_{I}^{-}\right),\qquad0\leq C_{r}\leq1\label{eq:4.3}
\end{equation}
Here, $x_{1}\left(t\right)$ and $x_{2}\left(t\right)$ are the principal
horizontal displacements of the center of mass of the superstructure
relative to the base of the frame in the unstepped and stepped phases
of motion, respectively, as shown in Fig. \ref{fig:Stepping-pier-.},
while coordinates $x_{3}\left(t\right)$ and $x_{4}\left(t\right)$
represent the principal vertical displacements of the apex of the
frame in the unstepped and stepped phases of motion. Also, $\xi_{1}$
is the damping factor for the assumed ``hysteretic'' damping (see
discussion in \cite{Beck-1}), $x_{c}$ is the displacement corresponding
to transition between the unstepped and stepped phases of lateral
motion, which is given in \cite{Beck-1}, and $C_{r}$ is the coefficient
of restitution amounting to energy loss at times before ($t_{I}^{-}$),
and after impact ($t_{I}^{+}$); $\omega_{1},\ldots,\omega_{4},P_{1},P_{2},\gamma_{1},\gamma_{2}$
in \eqref{eq:4.1} and \eqref{eq:4.2} are defined in \cite{Beck-1}.
Dependence on time in \eqref{eq:4.1} and \eqref{eq:4.2} is implicit;
single and double overdots denote the first and second derivatives
with respect to time. Note that the system of nonlinear differential
equations in \eqref{eq:4.1} and \eqref{eq:4.2} is stiff and coupled
befitting implicit numerical methods of solution.

The stepping frame problem generally involves both horizontal and
vertical degrees of freedom. The vertical motion is excited because
of large displacements (deviations from tangent to the displacement
curvature), pier touchdown, and vertical ground motion; but \emph{vertical
oscillations are possible even for purely horizontal excitation} \cite{Psycharis}.
In absence of vertical excitation however, and when calculating lateral
displacements as the primary measure of response, the influence of
vertical motion on the lateral displacements may be ignored \cite{Beck-1}.
For rigid blocks using a ``large'' ensemble of ground motions in
a certain probabilistic framework, the influence of the vertical component
of ground motion on rocking response is shown to be statistically
insignificant \cite{Vamvatsikos-2}. Still, for flexible portal frames
these effects call for further investigation that should include energy
exchange between lateral and vertical motion and different damping
mechanisms because of continual strain changes during touchdown \cite{Beck-1}.

\subsubsection{Stepping Portal Frame Dynamics}

\begin{figure}
\begin{centering}
\includegraphics{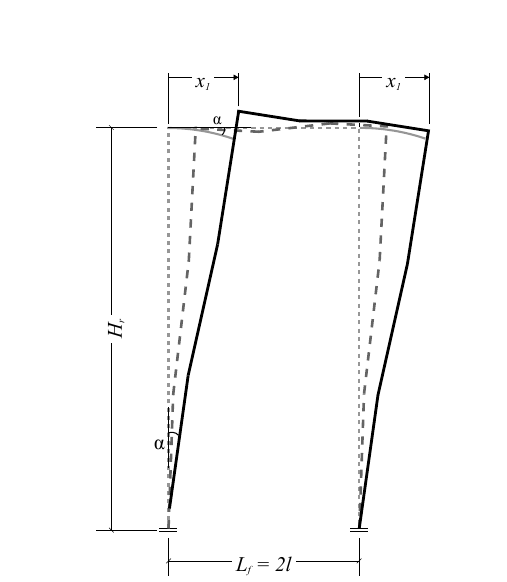}
\par\end{centering}
\caption{\label{fig:Portal-Frame-at}A stepping portal frame.}
\end{figure}

For portal frames such as the frame shown in Fig. \ref{fig:Portal-Frame-at},
$\theta=0$ and the \emph{unstepped phase of motion }is simplified
to a set of two second-order ordinary differential equations where
dependency on $t$ is implicit:
\begin{equation}
\left\{ \begin{array}{c}
\ddot{x}_{1}+2\xi_{1}\omega_{1}\dot{x}_{1}+\omega_{1}^{2}x_{1}=-P_{1}\ddot{x}_{g}\\
\ddot{x}_{3}+\omega_{3}^{2}x_{3}=0.
\end{array}\right.\qquad\qquad\qquad\begin{array}{c}
\mathrm{if}\end{array}\left|x_{1}\right|\leq x_{c}\label{eq:4.4}
\end{equation}
and similarly the \emph{stepped phase} in \eqref{eq:4.5} in which
the equation for stepping response $x_{2}$ is coupled and nonlinear:

\begin{equation}
\left\{ \begin{array}{c}
\ddot{x}_{2}+2\xi_{2}\omega_{2}\dot{x}_{2}-\omega_{2}^{2}x_{1}+\mathrm{sign}\left(x_{1}\right)\gamma_{1}=-P_{2}\ddot{x}_{g}\\
\ddot{x}_{4}+\omega_{4}^{2}x_{4}=0.
\end{array}\right.\:\begin{array}{c}
\mathrm{if}\end{array}\left|x_{1}\right|>x_{c}\label{eq:4.5}
\end{equation}
The equation of motion for the stepped phase is nonlinear due to sudden
sign change of the restoring force at the onset of uplift. Note the
negative ``stiffness'' term in \eqref{eq:4.5}, which depends on
the mass and height of the frame in a similar way to the inverted
pendulum, and discontinuous damping at the inception of a stepping
phase (see §\ref{subsec:The-Effects-of}). The parameters in \eqref{eq:4.4}
and \eqref{eq:4.5} are similar to those given for the A-frame:

\begin{equation}
\begin{array}{c}
\omega_{1}^{2}=\dfrac{2K_{b}\cdot g}{W_{S}+\frac{2}{3}W_{col.}},\;P_{1}=\dfrac{W_{S}+W_{col.}}{W_{S}+\frac{2}{3}W_{col.}}\\
\omega_{3}^{2}=\dfrac{2K_{a}\cdot g}{W_{S}+\frac{2}{3}W_{col.}}
\end{array}\label{eq:4.6}
\end{equation}
and:

\begin{equation}
\begin{array}{c}
\omega_{2}^{2}=\dfrac{W_{S}+W_{col.}}{W_{S}+\frac{5}{9}W_{col.}}\cdot\dfrac{g}{H_{r}},\;\gamma_{1}=\dfrac{W_{S}+2W_{col.}}{W_{S}+\frac{5}{9}W_{col.}}\cdot\dfrac{g}{H_{r}}.l,\;P_{2}=\dfrac{W_{S}+W_{col.}}{W_{S}+\frac{5}{9}W_{col.}}\\
\omega_{4}^{2}=\dfrac{K_{a}\cdot g}{W_{S}+\frac{4}{3}W_{col.}}\;\gamma_{2}=\dfrac{W_{S}+2W_{col.}}{W_{S}+\frac{5}{9}W_{col.}}\cdot g\cdot\tan\left(\theta\right)\cdot\sin\left(\theta\right)
\end{array}\label{eq:4.7}
\end{equation}
with $K_{a}$ and $K_{b}$ defined as the resultant axial and bending
stiffness of each leg. The $W_{S},W_{col.}$ and $\theta$ quantities
are defined in Fig. \ref{fig:Stepping-pier-.}. 

The transition from unstepped to stepped phase of response occurs
when lateral displacement $x_{1}$ in \eqref{eq:4.4} becomes large
enough to reduce the compression force in one of the footings to zero.
For a portal frame, the axial force induced by the \emph{dominant}
lateral displacement $x_{1}$ in the footing is $K_{a}x_{1}\sin\left(x_{1}/H_{r}\right)\cos\left(x_{1}/H_{r}\right)$,
resisted by half of the weight of the superstructures (for two columns)
and the column's weight. Hence, for the frame to remain in unstepped
phase, the axial compression due to the weight of the structure should
remain larger than the tensile force due to lateral displacement:

\begin{equation}
\frac{W_{S}}{2}+W_{col.}-\dfrac{1}{2}K_{a}\cdot x_{1}\cdot\sin\left(2\dfrac{x_{1}}{H_{r}}\right)>0\label{eq:4.8}
\end{equation}
The smallest value of $x_{1}$ at which condition \eqref{eq:4.8}
is violated corresponds to the critical value of lateral displacement,
$x_{c}$, at the onset of stepping, which can be expressed in terms
of the axial force in the footings as: 

\begin{equation}
\frac{W_{S}}{2}+W_{col.}=P_{t}\triangleq\frac{1}{2}K_{a}x_{c}\sin\left(\frac{2x_{c}}{H_{r}}\right)\label{eq:4.10}
\end{equation}
Hence, the dominant lateral displacement of a stepping portal frame
is described by a set of two ordinary second-order differential equations:

\begin{equation}
\begin{array}{cc}
\ddot{x}_{1}+2\xi_{1}\omega_{1}\dot{x}_{1}+\omega_{1}^{2}x_{1}=-P_{1}\ddot{x}_{g}\qquad & \left|x_{1}\right|<x_{c}\end{array}\label{eq:4.11}
\end{equation}

\begin{equation}
\begin{array}{cc}
\ddot{x}_{1}+2\xi_{2}\omega_{2}\dot{x}_{1}-\omega_{2}^{2}x_{1}+\mathrm{sign}\left(x_{1}\right)\gamma_{1}=-P_{2}\ddot{x}_{g}\qquad & \left|x_{1}\right|\geq x_{c}\end{array}\label{eq:4.12}
\end{equation}

\subsubsection{\label{subsec:The-Effects-of}The Effects of Damping}

The stepping response is non-conservative. The reduction in kinetic
energy during an inelastic impact is \cite{Housner-2}:
\begin{equation}
C_{r}^{2}=\left[\frac{\dot{x}_{2}\left(t_{I}^{+}\right)}{\dot{x}_{1}\left(t_{I}^{-}\right)}\right]^{2}\label{eq:4.13}
\end{equation}
using Eq. \eqref{eq:4.3}. From conservation of momentum:

\begin{equation}
I_{o}\dot{x}_{1}\left(t_{I}^{-}\right)-\left[\left(\bar{m}_{col.}H_{r}L_{F}^{2}\dot{x}_{1}\left(t_{I}^{-}\right)\right)+\left(\frac{1}{3}\bar{m}_{s}L_{F}^{3}\dot{x}_{1}\left(t_{I}^{-}\right)\right)\right]=I_{o}\dot{x}_{2}\left(t_{I}^{+}\right)\label{eq:4.14}
\end{equation}
so:

\begin{equation}
C_{r}=1-\frac{\bar{m}_{col.}H_{r}L_{F}^{2}+\frac{1}{3}\bar{m}_{s}L_{F}^{3}}{I_{o}}\label{eq:4.15}
\end{equation}
where the moment of inertia $I_{o}$ is given by:

\begin{equation}
I_{o}=\left(\frac{2}{3}\bar{m}_{col.}H_{r}+\bar{m}_{s}L_{F}\right)H_{r}^{2}\label{eq:4.16}
\end{equation}
in which $\bar{m}_{col.}=\frac{W_{col.}}{gH_{r}}$ and $\bar{m}_{s}=\frac{W_{S}}{gL_{F}}$
are the distributed column and superstructure unit masses. For $W_{S}=c_{1}W_{col.}$
and $L_{F}=c_{2}H_{r}$, where $1<c_{1}$ and $c_{2}<1$ are positive
real numbers, the coefficient of restitution $C_{r}=1-\left(\frac{3+c_{1}}{2+3c_{1}}\right)c_{2}^{2}$
has its relative maximum near $c_{2}\rightarrow0$ (i.e., for elastic
collision without loss of kinetic energy for tall frames, $C_{r}\rightarrow1$);
and relative minimum near $c_{1}\rightarrow1$ and $c_{2}\rightarrow1$
($C_{r}\rightarrow0$ for square frames). The equivalent viscous damping
ratio for equivalent linearization of a single degree of freedom oscillator
is shown to be a function of $C_{r}$ in \cite{Priestley-2}: 

\begin{equation}
\xi\backsimeq0.48\left(1-C_{r}^{2}\right)\label{eq:4.17}
\end{equation}

The assumption of inelastic impact (i.e., no bouncing) has been investigated
in free rocking response of prismatic rigid blocks for different ranges
of height to width ratio \cite{Pellegrino} and it has been found
generally adequate for slender blocks that are more susceptible to
rocking. It has been noted in other studies that if the assumption
of inelastic impact is not justified then some adjustment for equivalent
viscous damping may be required to account for return of the energy
of impact into the rocking system \cite{Priestley-1}.  

\subsection{General Forced Periodic Solution}

We aim to understand the dynamic response of the system described
by \eqref{eq:4.12}. The equation is nonlinear and separation of the
homogeneous free vibration and the particular forced oscillation solutions
is not possible, making influence of the initial conditions on response
longer lasting. 

Assume that $x_{1}\left(t\right)$ is a periodic solution with period
$\nicefrac{2\pi}{\Omega}$ when the system is subjected to a harmonic
forcing function:

\begin{equation}
\ddot{x}_{1}\left(t\right)-\omega_{2}^{2}x_{1}\left(t\right)+\mathrm{sign}\left[x_{1}\left(t\right)\right]\gamma_{1}=\bar{A}\cos\bar{\omega}t\label{eq:5.1}
\end{equation}
This is relevant, as will be shown later, in the solution of \eqref{eq:4.12}
subject to seismic excitation. We write $x_{1}\left(t\right)$ as
a Fourier series for all $t$:

\begin{equation}
x_{1}\left(t\right)=a_{0}+a_{1}\cos\Omega t+b_{1}\sin\Omega t+a_{2}\cos2\Omega t+b_{2}\sin2\Omega t+\cdots\label{eq:5.2}
\end{equation}
and substitute into \eqref{eq:5.1}:

\begin{equation}
A_{0}+A_{1}\cos\Omega t+B_{1}\sin\Omega t+A_{2}\cos2\Omega t+\cdots=\bar{A}\cos\bar{\omega}t\label{eq:5.3}
\end{equation}
where the $A_{i}$'s and $B_{i}$'s are functions of the $a_{i}$
and $b_{i}$, giving an infinite set of equations for them with $\Omega=\overline{\omega}$
for a solution \cite{Jordan-1}. Assuming that the response is mainly
dominated by a few harmonics, justified by a predominantly single-mode
of vibration of the stepping portal frame, one can use the truncated
Fourier series for $x_{1}\left(t\right)$:

\begin{equation}
x_{1}\left(t\right)=a_{0}+a_{1}\cos\overline{\omega}t+b_{1}\sin\overline{\omega}t\label{eq:5.4}
\end{equation}
in \eqref{eq:5.1} and by matching the coefficients of $\sin\overline{\omega}t$
and $\cos\overline{\omega}t$, solve for the amplitude of response:

\begin{equation}
a_{0}=\mathrm{sign}\left(x_{1}\right)\frac{\gamma_{1}}{\omega_{2}^{2}}\label{eq:5.5}
\end{equation}

\begin{equation}
b_{1}=0,\qquad a_{1}=\frac{-\bar{A}}{\overline{\omega}^{2}+\omega_{2}^{2}}\label{eq:5.6}
\end{equation}
showing that the approximate response is periodic in $t$ with $T=\nicefrac{2\pi}{\overline{\omega}}$
and out-of-phase with the excitation. 

\begin{figure}
\begin{centering}
\includegraphics[scale=0.7]{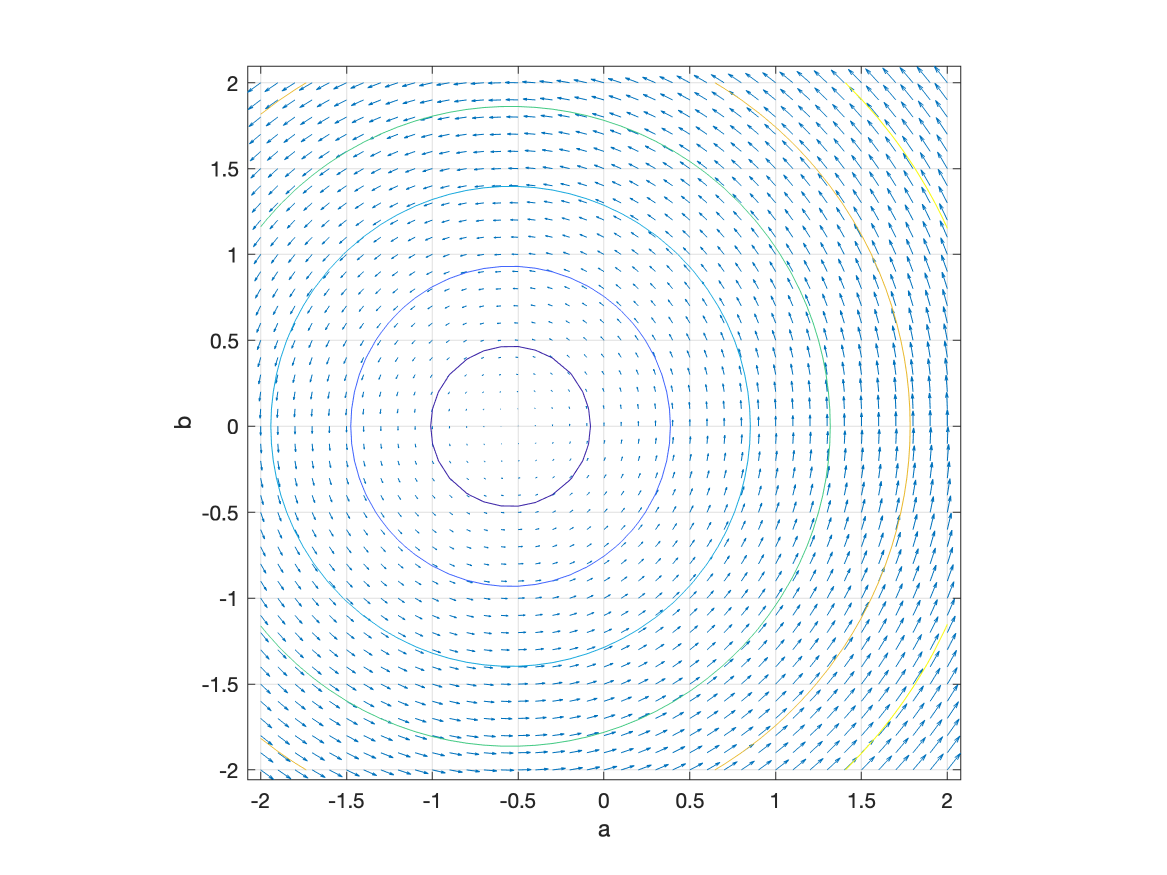}
\par\end{centering}
\caption{\label{fig:Phase-diagram-of}Phase diagram of \eqref{eq:5.10} for
$\omega_{2}=0.632$ rad/s, $\overline{\omega}=4.200$ rad/s, and $\bar{A}=9.820$
$\mathrm{m\cdot s^{-2}}$.}

\end{figure}

The stability of the solution is studied next by assuming that the
coefficients of response vary slowly as functions of time and checking
whether transient states attract or repel from the periodic solution
in \eqref{eq:5.4}:

\begin{equation}
x_{1}\left(t\right)=a_{0}\left(t\right)+a_{1}\left(t\right)\cos\overline{\omega}t+b_{1}\left(t\right)\sin\overline{\omega}t\label{eq:5.7}
\end{equation}
where $a_{0},\,a_{1},\,b_{2}$ are slowly varying compared to $\cos\overline{\omega}t$
and $\sin\overline{\omega}t$ and so their second derivatives may
be neglected: 

\begin{equation}
\dot{x}_{1}\left(t\right)=\dot{a}_{0}+\left(\dot{a}_{1}+b_{1}\overline{\omega}\right)\cos\overline{\omega}t+\left(\dot{b}_{1}-a_{1}\overline{\omega}\right)\sin\overline{\omega}t\label{eq:5.8}
\end{equation}

\begin{equation}
\ddot{x}_{1}\left(t\right)\simeq\left(2\dot{b}_{1}\overline{\omega}-a_{1}\overline{\omega}^{2}\right)\cos\overline{\omega}t-\left(2\dot{a}_{1}\overline{\omega}+b_{1}\overline{\omega}^{2}\right)\sin\overline{\omega}t\label{eq:5.9}
\end{equation}
Substituting in \eqref{eq:5.1} gives a system of autonomous equations
for $a_{1}$ and $b_{1}$:

\begin{equation}
\left\{ \begin{array}{c}
-2\dot{a}_{1}\overline{\omega}-b_{1}\overline{\omega}^{2}-\omega_{2}^{2}b_{1}=0\rightarrow\dot{a}_{1}=\frac{\overline{\omega}^{2}+\omega_{2}^{2}}{-2\overline{\omega}}b_{1}\triangleq\mathcal{A}\left(b_{1}\right)\\
2\dot{b}_{1}\overline{\omega}-a_{1}\overline{\omega}^{2}-\omega_{2}^{2}a_{1}=0\rightarrow\dot{b}_{1}=\frac{\bar{A}+a_{1}\left(\overline{\omega}^{2}+\omega_{2}^{2}\right)}{2\overline{\omega}}\triangleq\mathcal{B}\left(a_{1}\right)
\end{array}\right.\label{eq:5.10}
\end{equation}
subject to initial conditions:

\begin{equation}
\left\{ \begin{array}{c}
a_{1}\left(0\right)=-a_{0}+x_{1}\left(0\right)\\
b_{1}\left(0\right)=\frac{\dot{x}\left(0\right)}{\bar{\omega}}
\end{array}\right.\label{eq:5.11}
\end{equation}
where fixed (equilibrium) points of the coefficients of response $\left(a_{1}^{*},b_{1}^{*}\right)=\left(-\frac{\bar{A}}{\overline{\omega}^{2}+\omega_{2}^{2}},0\right)$
are obtained from $\mathcal{A}\left(b_{1}\right)=\mathcal{B}\left(a_{1}\right)=0$.
Note that $a_{1}^{*}$ is a nonpositive real number typically in $\left[-10^{2},0\right]$
and the fixed point is a center as shown by the phase diagram in Fig.
\ref{fig:Phase-diagram-of}.

Hence, the solution to \eqref{eq:5.1} is expressed as:

\begin{equation}
x_{1}\left(t\right)=\frac{\mathrm{sign}\left(x_{1}\right)\gamma_{1}}{\omega_{2}^{2}}-\frac{\bar{A}}{\overline{\omega}^{2}+\omega_{2}^{2}}\cos\bar{\omega}t\label{eq:5.12}
\end{equation}
where $\bar{A}\triangleq\left|\ddot{x}_{g}\right|_{max}P_{2}$. Substituting
for $\gamma_{1}$, $\omega_{2}^{2}$, and $\bar{A}$:

\begin{equation}
x_{1}\left(t\right)\approx\frac{L_{F}}{2}\mathrm{sign}\left(x_{1}\right)-\frac{\left|\ddot{x}_{g}\right|_{max}}{\overline{\omega}^{2}+\frac{g}{H_{r}}}\cos\bar{\omega}t\label{eq:5.13}
\end{equation}
Note that for short frames ($H_{r}\rightarrow0$), the oscillating
term is insignificant and $\left|x_{1}\right|\rightarrow\pm\nicefrac{L_{F}}{2}$.
For tall frames ($H_{r}\rightarrow\infty$) under long-period excitation,
the response increases rapidly $\left|x_{1}\right|\rightarrow\nicefrac{L_{F}}{2}+\nicefrac{\left|\ddot{x}_{g}\right|_{max}}{\bar{\omega}^{2}}$
whereas high frequency excitation produces little dynamic response
when the frame is tall. 
\begin{rem}
It is common in earthquake engineering to represent ground motion
as a Fourier series, assuming periodic loading. In that sense, \eqref{eq:5.1}
is the equation of motion of the stepping frames subject to a ground
motion in which only one harmonic amplitude coefficient is significant. 
\end{rem}

\subsubsection{Example 3:\label{subsec:Example-3} Harmonic Excitation}

We consider a flexible frame similar to §\ref{subsec:Example-2} subject
to harmonic excitation $A_{m}\cos\bar{\omega}t$. For uniform rocking
blocks under harmonic excitation, sensitivity of solutions to initial
conditions is well-established \cite{Bruhn,Hogan-3,Hogan-1,Hogan-2,Spanos-1}.
It has been noted that ``\emph{the simple model of Housner (1963)
is shown to possess extremely complicated dynamics, including chaos}''
\cite{Hogan-2} and ``\emph{highly counter-intuitive observations}''
of a rocking block ``\emph{during forced motion beyond the point
at which it would topple if it were not being forced}''\footnote{Implication for standing precarious rocks, see Fig. \ref{fig:Example-of-precarious}.}
\cite{Hogan-2}. We are interested in finding out if such sensitivity
also exists in the stepping response of flexible frames. Aperiodic
long-term behavior would have to be considered carefully when establishing
practical limits of stability under random external perturbations. 

\begin{figure}
\begin{centering}
\includegraphics[scale=0.7]{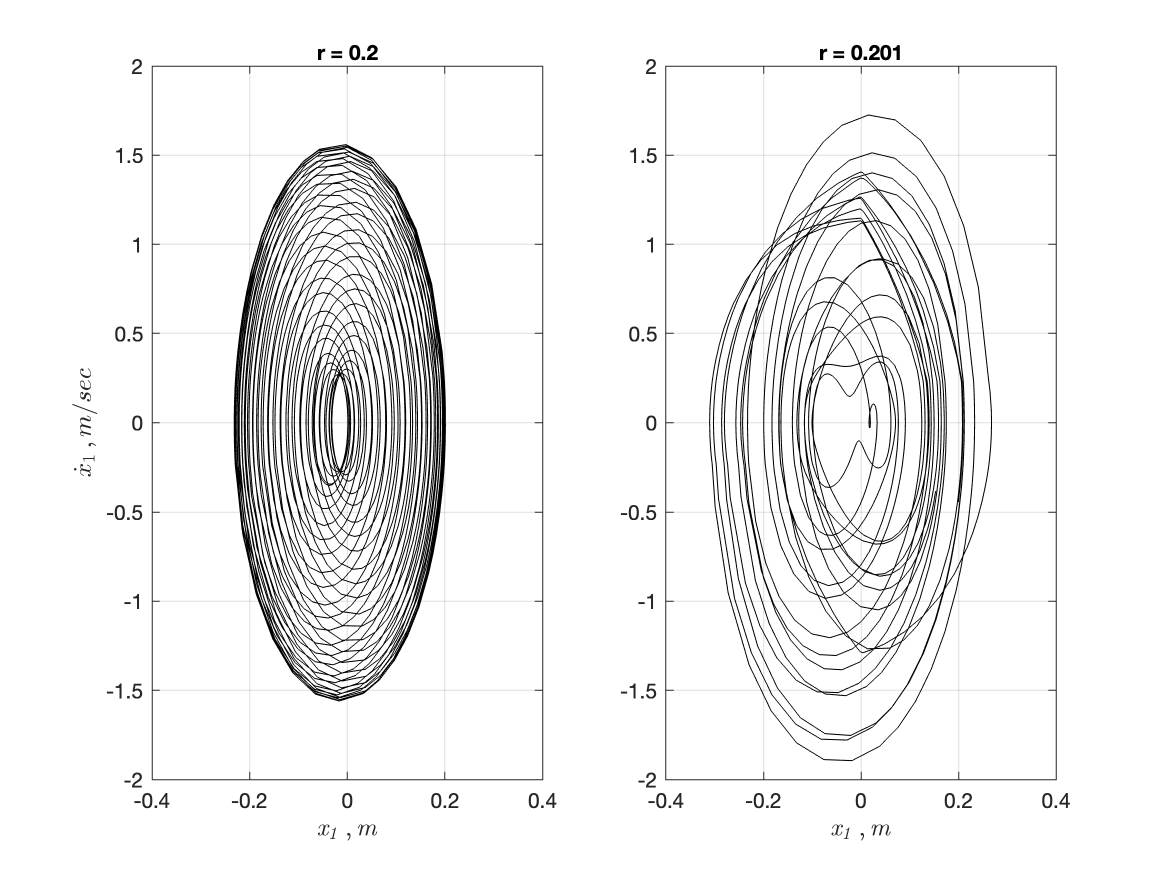}
\par\end{centering}
\caption{\label{fig:Transition-of-phase}Transition of phase portrait at initiation
of rocking. }

\end{figure}

\begin{figure}
\begin{centering}
\includegraphics[scale=0.7]{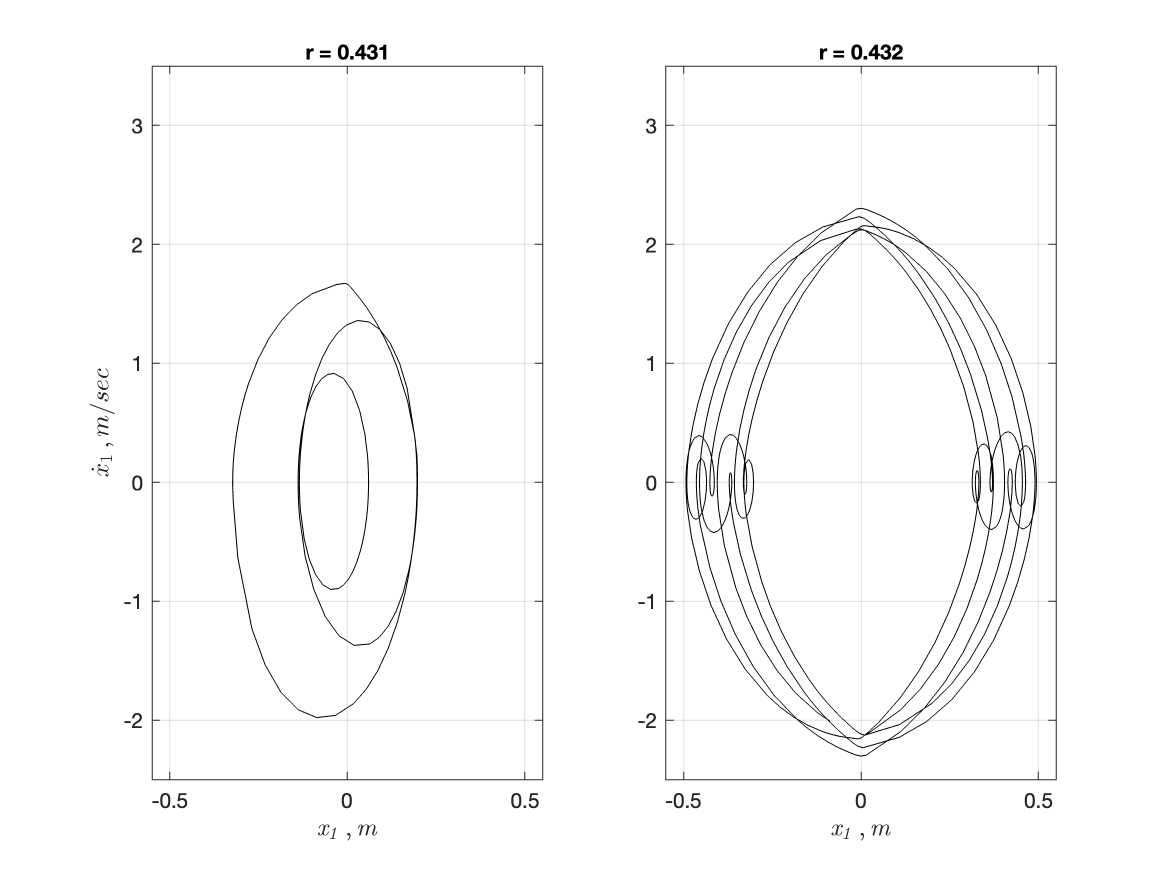}
\par\end{centering}
\caption{\label{fig:Bifurecation-of-response}Bifurcation of response from
asymmetric $\left(1,1\right)$ to rotationally symmetric $\left(5,15\right)$
orbits at $r=0.431.$}
\end{figure}

Shown in Figs. \eqref{fig:Transition-of-phase} and \eqref{fig:Bifurecation-of-response}
are orbits of \eqref{eq:4.11} and \eqref{eq:4.12} numerically solved
starting from $x_{1}\left(0\right)=\dot{x}_{1}\left(0\right)=0$ after
allowing the system to evolve over a long time and disregarding the
initial transient response \cite{Strogatz-1}. The parameter set is
$H_{r}=29.944$ m, $L_{F}=8.839$ m, $W_{T}=2.807\times10^{6}$ N,
$W_{s}=1.472\times10^{6}$ N, $W_{col.}=\frac{1}{2}\left(W_{T}-W_{s}\right)$,
$K_{a}=1.159\times10^{9}$ N/m, $K_{b}=5.019\times10^{6}$ N/m, $\xi_{1}=0.05$,
$\bar{\omega}=7.854$ rad/s, and $\bar{A}\triangleq r\cdot g$ where
$g=9.81\mathrm{m\cdot s^{-2}}$ and the values of $r$ are shown above
the panels in the figures. A variable-step, variable-order algorithm
for stiff differential equations (ODE15s) is used for numerical integration
\cite{Matlab}. 

For a rigid block with uniform mass, the onset of rocking will be
at $r=\nicefrac{L_{f}}{H_{r}}=0.2952$ per West's formula \cite{Milne}
under static conditions\footnote{West's equation has been used in the past to estimate intensity of
ground motion from the sizes of overturned tomb stones in Japan \cite{Otani}.}. For a rigid frame, the corresponding value is:

\begin{equation}
r\geq\frac{L_{f}}{H_{r}}\left(\frac{\frac{W_{S}}{2}+W_{col.}}{W_{S}+W_{col.}}\right)\label{eq:5.14}
\end{equation}
or $r\geq0.1936$. Fig. \ref{fig:Transition-of-phase} shows the transition
from unstepped to stepped phase occurring at $r=0.201$ for the flexible
frame. Stable cyclic response under forced vibration beyond the point
of static stability is plausible for flexible frames.

Fig. \ref{fig:Bifurecation-of-response} presents an instance of symmetry
breaking and bifurcation in forced harmonic response of the stepping
flexible frame. We note discontinuity of trajectories at $x_{1}=0$
(``pinching'') and speculate that this is an inherent feature of
rocking dynamics as it is also shown in unforced rigid block dynamics
by Hogan \cite{Hogan-1}. 

The saddle-node bifurcations appear at $r>0.0399$ calculated from
\eqref{eq:5.15} \cite{Bruhn}:

\begin{equation}
r=\frac{\left(1+\bar{\omega}^{2}\right)\left(1-\sqrt{\rho}\right)\left[\cosh\left(\frac{n\pi}{\bar{\omega}}\right)-1\right]}{\sqrt{\bar{\omega}^{2}\left(1-\sqrt{\rho}\right)^{2}\sinh^{2}\left(\frac{n\pi}{\bar{\omega}}\right)+\left(1+\sqrt{\rho}\right)^{2}\left[\cosh\left(\frac{n\pi}{\bar{\omega}}\right)+1\right]^{2}}}\label{eq:5.15}
\end{equation}

\begin{figure}
\begin{centering}
\includegraphics[scale=0.7]{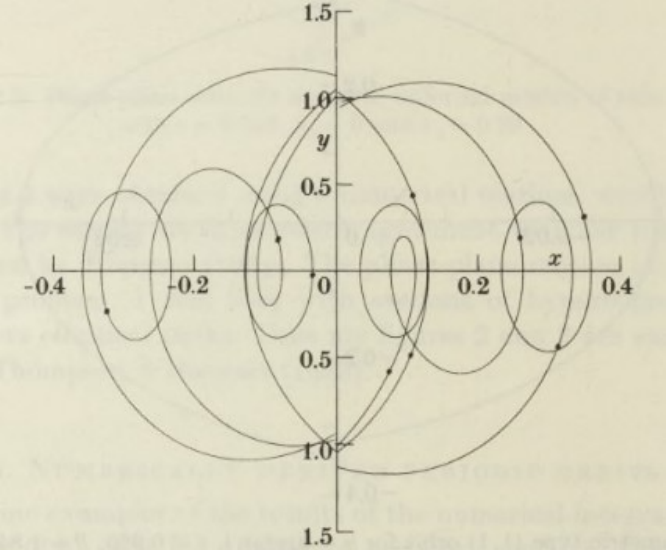}
\par\end{centering}
\caption{\label{fig:Symmetric-orbits-for}Example of $\left(3,9\right)$ symmetric
orbits in response of a rigid rocking block in \cite{Hogan-1}. }

\end{figure}

Note the bifurcation in Fig. \ref{fig:Bifurecation-of-response} is
similar to those of rigid blocks shown in Fig. \ref{fig:Symmetric-orbits-for}.

\begin{rem}
The complexity of response under a simple forcing function has been
observed in other studies but rarely in the context of chaotic dynamics
of \eqref{eq:4.4} and \eqref{eq:4.5} or in earthquake engineering.
Numerical simulations have indicated heightened sensitivity to changes
in size and details of the forcing function with no systematic trends
\cite{Penzien-1}. Implications for this will be discussed next. 
\end{rem}

\subsection{Seismic Response}

We are interested in extending, by numerical experiments, the observations
of the complicated dynamics of rigid blocks studies by others to the
response of stepping flexible frames, and learning about the implications
that they may have on establishing a rational method of probabilistic
seismic performance assessment of stepping flexible frames. Given
the preceding intricacies in the dynamics of stepping bodies (§\ref{subsec:Example-3}),
the influence of the assumptions of quasi-dynamic response analysis
(§\ref{sec:Equivalent-Linearization-of}), and the consequences of
overturning on safety and functionality of engineered structures,
it is prudent to go beyond simple methods of preliminary design of
flexible stepping frames to better understand their nonlinear dynamic
behavior. This can be achieved by using time-stepping analyses under
an ensemble of site-appropriate ground motion time histories. This
section is an example of such analyses.

\subsubsection{Moving Resonance}

\label{Moving-resonance-}Moving resonance \cite{Beck-2,Wen-1} is
an inherent feature of dynamic response of softening nonlinear oscillators
under nonstationary stochastic excitation and it occurs when changes
in system frequencies track the shift of dominant frequencies in the
excitation with increasing amplitudes. While resonance describes exponential
growth of response in linear oscillators when their natural frequency
of vibration is at or close to the excitation frequency, the moving
resonance portrays a similar phenomenon in a nonlinear oscillator
when the evolution of natural frequency of vibration of the oscillator
follows that of the nonstationary excitation. As in inverted pendula
in which the period of oscillation grows with the amplitude of sway,
the rocking period of vibration elongates logarithmically with the
amplitude of displacements (see, e.g., Fig. 2 in \cite{Housner-2}
for rigid blocks or Fig. 3 in \cite{Beck-1} for stepping frames).
On the other hand, the temporal shift of frequency content towards
lower frequencies is rather common in many earthquake ground motions,
as observed in accelerograms. It has been shown that the effect of
temporal nonstationairty in the frequency content of excitation on
response of nonlinear oscillators can be severe, causing amplification
of response by a factor as high as three \cite{Beck-2}. We investigate
the phenomenon and propose an approach that could be useful in addressing
it during design.

We examine the following equivalent linear model involving the equation
of motion for a time-varying linear oscillator oscillator subject
to excitation $\ddot{x}_{g}(t)$ with temporal nonstationarities \cite{Beck-2}:

\begin{equation}
\underline{\dot{x}}(t)=\mathbf{A}(t)\underline{x}(t)+\underline{g}(t)\label{eq:6.1}
\end{equation}
\textcolor{black}{where}

\begin{equation}
\underline{x}(t)=\left\{ \begin{array}{c}
x(t)\\
\dot{x}(t)
\end{array}\right\} ,\:\mathbf{A}(t)=\left[\begin{array}{cc}
0 & 1\\
-\omega^{2}(t) & -2\xi(t)\omega(t)
\end{array}\right],\:\underline{g}(t)=\left\{ \begin{array}{c}
0\\
\ddot{x}_{g}(t)
\end{array}\right\} \label{eq:6.2}
\end{equation}
and where $\ddot{x}_{g}(t)$ is a nonstationary stochastic ground
motion process modeled by the output of two cascaded second-order
differential equations: 

\begin{equation}
\left\{ \begin{array}{c}
\ddot{y}+2\xi_{g}(t)\omega_{g}(t)\dot{y}+\omega_{g}^{2}(t)y=f_{g}(t)e(t)\\
\ddot{x}_{g}+2\omega_{c}\dot{x}_{g}+\omega_{c}^{2}x_{g}=\ddot{y}
\end{array}\right.\label{eq:6.3}
\end{equation}
In \eqref{eq:6.3}, $\xi_{g}(t)=\alpha_{g}(t)\cdot\omega_{g}^{-1}(t)$,
where $\omega_{g}(t)$ and $\alpha_{g}(t)$ are deterministic, slowly-varying,
functions that approximate the dominant ground acceleration frequency
and its bandwidth described below, $f_{g}(t)$ is also a deterministic
function and it represents the envelop function that modulates the
amplitudes of the ground motion acceleration process, $e(t)$ is a
Gaussian white-noise process with zero mean and identity auto-correlation
matrix, and $\omega_{c}$ is the corner frequency from the source
mechanism of the earthquake that generates the ground motion \cite{Brune-1,Heaton-1,Kanamori-1}. 

From the Liapunov differential matrix equation, the mean-square of
displacement response of the oscillator $q_{11}\left(t\right)=E\left[\left(x^{2}\left(t\right)\right)\right]$
subject to broadband process $\ddot{x}_{g}(t)$ is obtained from the
solution of the first-order differential equation \cite{Papadimitriou-1}:

\begin{equation}
\dot{q}_{11}\left(t\right)+2\left[\xi_{0}\omega_{0}+\frac{\dot{\omega}_{d}\left(t\right)}{2\omega_{d}\left(t\right)}\right]q_{11}\left(t\right)=\frac{I_{g}^{2}\left(t\right)R_{g}\left[\omega\left(t\right),\xi_{0}\omega_{0}\left|\underline{\theta}_{g}\left(t\right)\right.\right]}{2\omega^{2}\left(t\right)}\label{eq:6.4}
\end{equation}
where $\omega_{d}\left(t\right)=\left[\omega^{2}\left(t\right)-\xi_{0}^{2}\omega_{0}^{2}\right]^{\nicefrac{1}{2}}$
is the damped natural frequency, $\omega_{0}$ and $\xi_{0}$ are
the small-amplitude oscillator frequency and viscous damping ratio,
respectively, $I_{g}\left(t\right)=\sqrt{E\left[\ddot{x}_{g}(t)\right]}$
is the amplitude intensity of the ground motion process, and $R_{g}$
represents the time-varying power spectral density of the ground motion
acceleration described next for $\underline{\theta}_{g}(t)=\left\{ \omega_{g}(t),\alpha_{g}(t)\right\} $,
the estimated ground motion parameter vector: 

\begin{equation}
R_{g}\left[\omega\left(t\right),\xi_{0}\omega_{0}\left|\underline{\theta}_{g}\left(t\right)\right.\right]=2I_{c}\left(t\right)+2\xi_{0}\omega_{0}I_{s}\left(t\right)\label{eq:6.5}
\end{equation}
where

\begin{equation}
I_{c}\left(t\right)=\int_{0}^{\infty}\exp\left(-\xi_{0}\omega_{0}\tau\right)r\left[\tau\left|\underline{\theta}_{g}\left(t\right)\right.\right]\cos\left(\omega_{d}\left(t\right)\tau\right)d\tau\label{eq:6.6}
\end{equation}

\begin{equation}
I_{s}\left(t\right)=\frac{1}{\omega_{d}\left(t\right)}\int_{0}^{\infty}\exp\left(-\xi_{0}\omega_{0}\tau\right)r\left[\tau\left|\underline{\theta}_{g}\left(t\right)\right.\right]\sin\left(\omega_{d}\left(t\right)\tau\right)d\tau\label{eq:6.7}
\end{equation}
$r\left[\tau\left|\underline{\theta}_{g}(t)\right.\right]$ is the
auto-correlation function of $\ddot{x}_{g}(t)$ at time $\tau$ which
can be described analytically, allowing integration of \eqref{eq:6.6}
and \eqref{eq:6.7} \cite{Beck-2,Papadimitriou-1}:

\begin{multline}
I_{c}\left(t\right)=\frac{\left[\xi_{0}\omega_{0}+\alpha_{g}\left(t\right)\right]\left\{ \left[\xi_{0}\omega_{0}+\alpha_{g}\left(t\right)\right]^{2}+\omega_{d}^{2}\left(t\right)+\omega_{g}^{2}\left(t\right)-\alpha_{g}^{2}\left(t\right)\right\} }{\left[\omega^{2}\left(t\right)-\omega_{g}^{2}\left(t\right)\right]^{2}+4\omega\left(t\right)\omega_{g}\left(t\right)\left[\xi_{0}\omega_{0}+\alpha_{g}\left(t\right)\right]\left[\xi_{g}\left(t\right)\omega\left(t\right)+\xi\left(t\right)\omega_{g}\left(t\right)\right]}+\\
\frac{+\alpha_{g}\left(t\right)\left\{ \left[\xi_{0}\omega_{0}+\alpha_{g}\left(t\right)\right]^{2}-\omega_{d}^{2}\left(t\right)+\omega_{g}^{2}\left(t\right)-\alpha_{g}^{2}\left(t\right)\right\} }{\left[\omega^{2}\left(t\right)-\omega_{g}^{2}\left(t\right)\right]^{2}+4\omega\left(t\right)\omega_{g}\left(t\right)\left[\xi_{0}\omega_{0}+\alpha_{g}\left(t\right)\right]\left[\xi_{g}\left(t\right)\omega\left(t\right)+\xi\left(t\right)\omega_{g}\left(t\right)\right]}\label{eq:6.8}
\end{multline}

\begin{multline}
I_{s}\left(t\right)=\frac{\left[\xi_{0}\omega_{0}+\alpha_{g}\left(t\right)\right]^{2}\left[\omega_{d}^{2}\left(t\right)-\omega_{g}^{2}\left(t\right)+\alpha_{g}^{2}\left(t\right)\right]+2\left[\xi_{0}\omega_{0}+\alpha_{g}\left(t\right)\right]\alpha_{g}\left(t\right)}{\left[\omega^{2}\left(t\right)-\omega_{g}^{2}\left(t\right)\right]^{2}+4\omega\left(t\right)\omega_{g}\left(t\right)\left[\xi_{0}\omega_{0}+\alpha_{g}\left(t\right)\right]\left[\xi_{g}\left(t\right)\omega\left(t\right)+\xi\left(t\right)\omega_{g}\left(t\right)\right]}\label{eq:6.9}
\end{multline}

In this context, moving resonance occurs when the time-varying power
spectral density of the ground motion process, $R_{g}$, on the RHS
of \eqref{eq:6.4} grows as $\omega\left(t\right)\rightarrow\omega_{g}(t)$
over some time interval. The ground motion parameters $\omega_{g}(t)$
and $\alpha_{g}(t)$ in \eqref{eq:6.8} and \eqref{eq:6.9} are solely
geophysical and functions of $\omega_{p}$, $\omega_{s}$, and $\omega_{r}$,
the estimated dominant frequencies of the P, S, and surface waves
in the earthquake ground motion, and the estimated frequency bands
around the dominant P and surface wave frequencies, respectively: 

\begin{equation}
\omega_{g}\left(t\right)=\omega_{r}+\left(\omega_{p}-\omega_{r}\right)\left(\frac{\omega_{s}-\omega_{r}}{\omega_{p}-\omega_{r}}\right)^{\frac{t}{t_{max}}}\label{eq:6.10}
\end{equation}

\begin{equation}
\alpha_{g}\left(t\right)=\omega_{p}\xi_{p}+\left(\omega_{r}\xi_{r}-\omega_{p}\xi_{p}\right)\frac{t}{t_{dur}}\label{eq:6.11}
\end{equation}
where $t_{max}$ is the time of the maximum intensity of the ground
motion intensity, i.e., $t_{max}=\mathrm{argmax}\left[I_{g}\left(t\right)\right]$
and $t_{dur}$ is the time interval over which $I_{g}\left(t\right)$
is greater than a predefined percentage of $\max\left[I_{g}\left(t\right)\right]$
\cite{Beck-2}.

\begin{figure}
\begin{centering}
\includegraphics[scale=1.2]{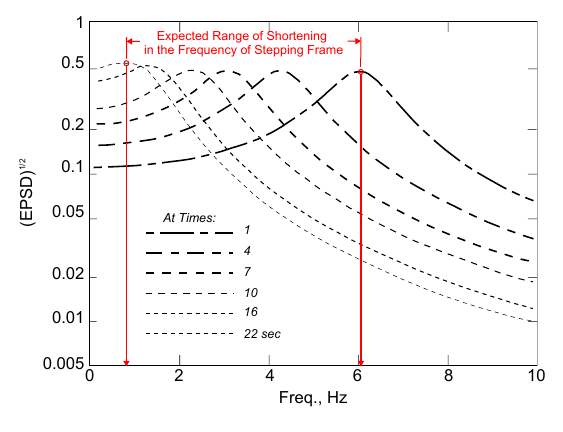}
\par\end{centering}
\caption{\label{fig:Forcing-function-selection}Forcing function selection
criterion (modified after \cite{Beck-2}). }

\end{figure}

As a matter of practical importance, proper selection of forcing functions
for safety assessment of stepping frames under seismic ground motions
is prudent. This is especially pertinent for sites located near major
active faults \cite{Hall-2}. Conditions\footnote{such as site, source and path of the seismic waves, and history of
slip on the causative fault.} that give rise to shifting oscillations towards lower frequencies
in the ground motion at the site may result in a detrimental evolutionary
power spectral density:

\begin{equation}
S\left(\omega,t\right)=\frac{4\alpha_{g}\left(t\right)\omega_{g}^{2}\left(t\right)I_{g}^{2}\left(t\right)}{\left[\omega^{2}-\omega_{g}^{2}\left(t\right)\right]^{2}+4\omega^{2}\alpha_{g}^{2}\left(t\right)}\label{eq:6.12}
\end{equation}
in conjunction with changes in the stepping response natural frequency
(e.g., see Fig. 2 in \cite{Housner-2} for rigid blocks and Fig. 3
in \cite{Beck-1} for stepping frames, §\ref{subsec:Discussion-of-period}),
and its small-amplitude bandwidth. To find conditions conductive to
moving resonance, we find ground motion time series with evolutionary
power spectral density at the site that match the corresponding peak-to-peak
frequencies with that of the stepping frame, as shown in Fig. \ref{fig:Forcing-function-selection}.

The amplification in response due to moving resonance, $\mathscr{A}$,
can be approximated by the ratio of peak time-varying power spectral
density of the ground motion acceleration as $t\rightarrow t_{max},\:\omega(t)\rightarrow\omega_{g}(t)$
to peak time-varying power spectral density of the ground motion at
small-amplitude oscillator frequency and viscous damping ratio as
$t\rightarrow0$: this is an approximation because \eqref{eq:6.4}
is a linear equation with time-dependent coefficients but for slowly
varying $\dot{\omega}_{d}(t)$ relative to $\omega_{d}(t)$: 

\begin{equation}
\mathscr{A}\triangleq\frac{R_{g}\left[\omega(t)\rightarrow\omega_{g}(t),\xi_{0}\omega_{0}\left|\underline{\theta}_{g}\left(t\right)\right.\right]}{R_{g}\left[\omega_{0},\xi_{0}\omega_{0}\left|\underline{\theta}_{g}\left(t\right)\right.\right]}\cdot\left[\frac{\omega_{0}}{\omega(t)}\right]^{2}\label{eq:6.13}
\end{equation}

From \eqref{eq:6.5} and \eqref{eq:6.8} to \eqref{eq:6.11}:

\begin{equation}
R_{g}\left[\omega(t)\rightarrow\omega_{g}(t),\xi_{0}\omega_{0}\left|\underline{\theta}_{g}\left(t\right)\right.\right]=\frac{10\alpha_{0}\alpha_{g}+4\omega_{g}^{2}+\alpha_{0}\left(\alpha_{0}+\alpha_{g}\right)\left(\alpha_{g}^{2}+\alpha_{0}^{2}\right)}{4\alpha_{g}\omega_{g}^{2}}\label{eq:6.14}
\end{equation}
where for $t\rightarrow t_{max}$, $\omega_{g}\rightarrow\omega_{s}$
and $\alpha_{g}\rightarrow\omega_{r}\xi_{r}$, and assuming negligible
small-amplitude viscous damping, $\xi_{0}=\alpha_{0}=0$:

\begin{equation}
R_{g}\left[\omega(t)\rightarrow\omega_{g}(t),\xi_{0}\omega_{0}\left|\underline{\theta}_{g}\left(t\right)\right.\right]=\left(\omega_{r}\xi_{r}\right)^{-1}\label{eq:6.15}
\end{equation}

On the other hand:
\begin{equation}
R_{g}\left[\omega_{0},\xi_{0}\omega_{0}\left|\underline{\theta}_{g}\left(t\right)\right.\right]=\frac{4\alpha_{g}\omega_{g}^{2}}{\left(\omega_{0}^{2}-\omega_{g}^{2}\right)^{2}+4\alpha_{g}^{2}\omega_{0}^{2}}\label{eq:6.16}
\end{equation}
where $t\rightarrow0$, $\omega_{g}\rightarrow\omega_{p}$ and $\alpha_{g}\rightarrow\omega_{p}\xi_{p}$,
therefore:

\begin{equation}
R_{g}\left[\omega_{0},\xi_{0}\omega_{0}\left|\underline{\theta}_{g}\left(t\right)\right.\right]=\frac{4\left(\omega_{p}\xi_{p}\right)\omega_{p}^{2}}{\left(\omega_{0}^{2}-\omega_{p}^{2}\right)^{2}+4\left(\omega_{p}\xi_{p}\right)^{2}\omega_{0}^{2}}\label{eq:6.17}
\end{equation}
and hence:

\begin{equation}
\mathscr{A}=\frac{\left(\omega_{0}^{2}-\omega_{p}^{2}\right)^{2}+4\left(\omega_{p}\xi_{p}\right)^{2}\omega_{0}^{2}}{4\left(\omega_{r}\xi_{r}\right)\left(\omega_{p}\xi_{p}\right)\omega_{p}^{2}}\cdot\left(\frac{\omega_{0}}{\omega_{s}}\right)^{2}\label{eq:6.18}
\end{equation}

\begin{figure}
\begin{centering}
\includegraphics[scale=0.7]{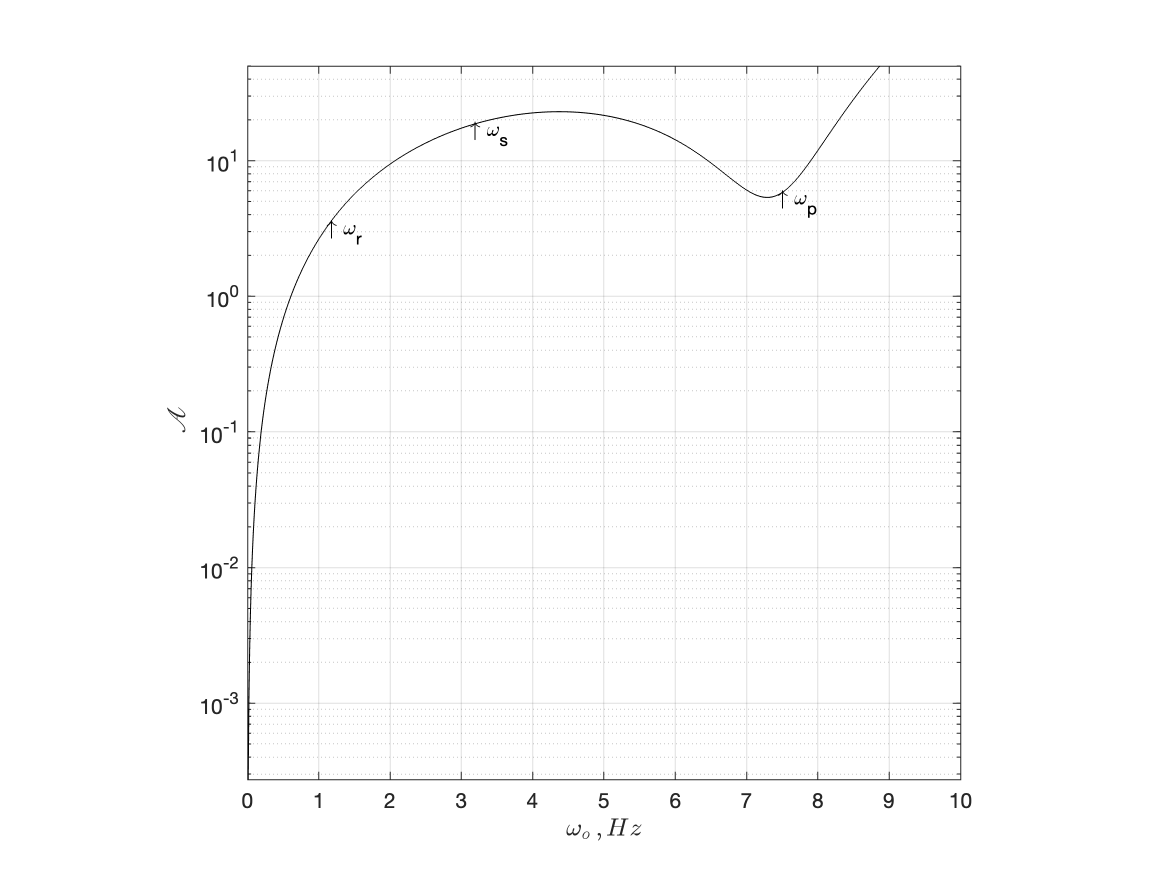}
\par\end{centering}
\caption{\label{fig:Variation-of-moving}Variation of the moving resonance
amplification factor as a function of small-amplitude oscillator frequency.}

\end{figure}

Note that for a small-amplitude oscillator frequency near the P-wave
dominant frequency, $\omega_{0}\thickapprox\omega_{p}$, the amplification
in response of the oscillator due to moving resonance is approximately:

\begin{equation}
\mathscr{A}\simeq\left(\frac{\omega_{p}\xi_{p}}{\omega_{r}\xi_{r}}\right)\cdot\left(\frac{\omega_{p}}{\omega_{s}}\right)^{2}\thickapprox\left(\frac{\omega_{p}}{\omega_{s}}\right)^{2}\label{eq:6.19}
\end{equation}
which is an entirely geophysical factor of the seismic wave characteristics
at the site. It depends on the material stress-strain constitutive
relationship, which in turn rests on the rigidity and bulk modulus
of the material \cite{Stein-1} with typical values ranging from two
to as high as ten for short-period body waves. For $\omega_{p}=7.43$
Hz, $\omega_{s}=3.12$ Hz, $\omega_{r}=1.11$ Hz, $\xi_{p}=0.096$,
and $\xi_{r}=0.655$ in the time-varying model of ground motion in
\cite{Beck-2}, the approximate amplification factor due to moving
resonance is 5.67. Eq. \eqref{eq:6.18} is shown graphically in Fig.
\ref{fig:Variation-of-moving} for $0.1\leq\omega_{0}\leq10$ Hz and
the aforementioned time-varying ground motion model parameters.

The fixed points of displacement in §\ref{subsec:Fixed-Points-of}
can be adjusted accordingly to include the effect of moving resonance
on quasi-dynamic response of stepping oscillators. The adjustments
in \eqref{eq:6.18} or \eqref{eq:6.19} are available from site-specific
spectral analysis of seismic time series or by using representative
phase velocities.\textcolor{red}{{} }

\subsubsection{Example 4:\label{subsec:Example-4} Seismic Excitation}

\begin{figure}
\begin{centering}
\includegraphics[scale=0.7]{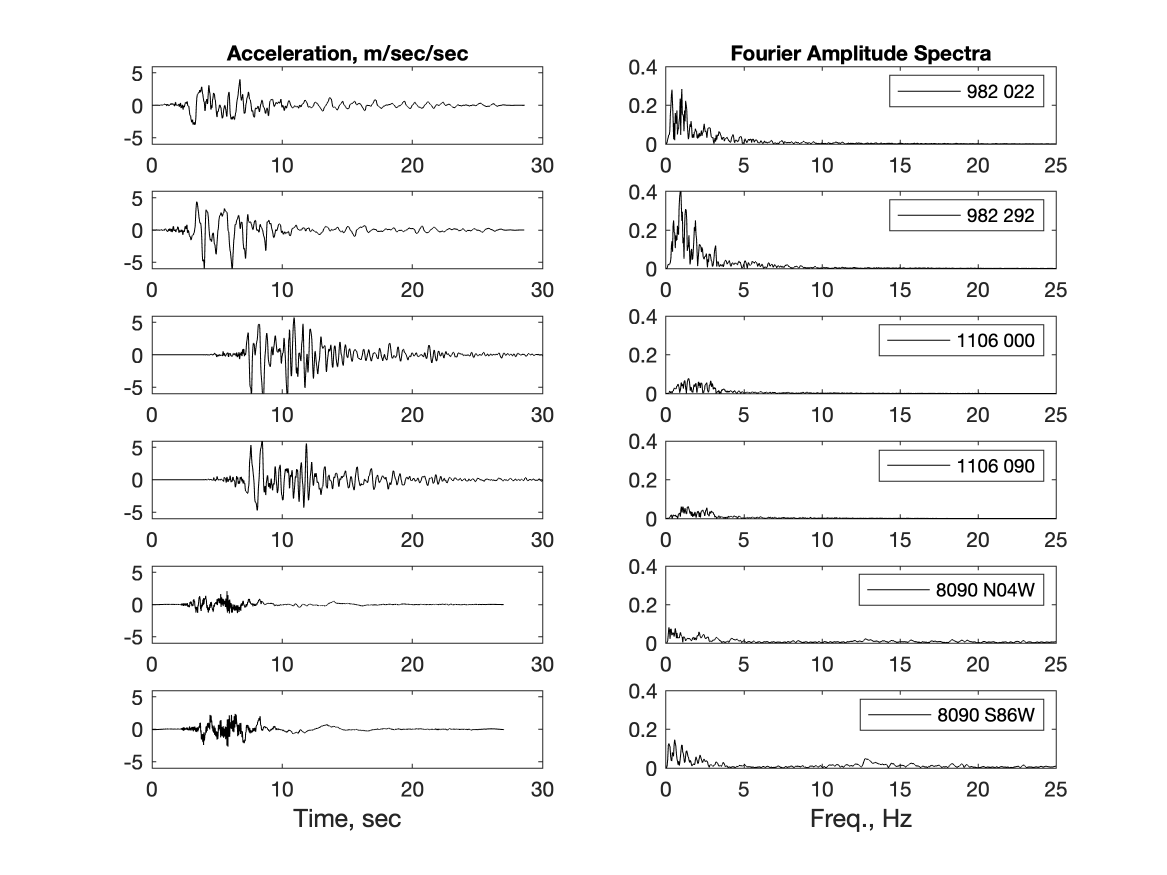}
\par\end{centering}
\caption{\label{fig:Input-motion-and}Recorded pairs of horizontal ground motions
and their Fourier amplitude spectra used in Example 4.}
\end{figure}

We consider a frame similar to Example 2 in §\ref{subsec:Example-2}
subject to a set of earthquake ground acceleration records listed
in Tables \ref{tab:The-events-considered} to \ref{tab:Ground-motion-characteristics}
and shown in Fig. \ref{fig:Input-motion-and}. Our investigation is
qualitative in nature to inform the paper's broader objective of seeking
insight into nonlinear dynamic response of flexible frames subject
to earthquake ground motions, and as such, we are not concerned with
different probabilities of hazard or whether the ensemble of ground
motion records chosen adhere to any specific character of a particular
site or whether they are suitable for any type of structure. No modifications
to the amplitudes or the frequency content of the time series were
made in this example. Two un-rotated horizontal components of the
ground motion acceleration time series were chosen from a set of 21540
records \cite{PEER-1} (see Record ID and Components in Table  \ref{tab:The-events-considered})
with no attempt at selecting the records on specific criteria except
to have their values of $S_{DS}$ and $S_{D1}$ as close to those
in Example 2 as possible. 

The three events, Northridge, California earthquake of Jan. 17, 1994;
Kobe, Japan earthquake of Jan. 16, 1995; and Christchurch, New Zealand
earthquake of Feb. 21, 2011 are all shallow crustal earthquakes. Table
\ref{tab:Event-characteristics-and} shows the moment magnitude, the
hypocentral depth, the epicentral distance, and the average shear
wave velocity of the top 30 m of strata at the site. These ground
motion records correspond to shaking at site classes described by
``loose sand or medium stiff clay'' to ``dense sand or very stiff
clay'' strata \cite{AASHTO-2,ASCE-1}. 

\begin{table}
\caption{\label{tab:The-events-considered}Seismic events considered in Example
4.}

\centering{}%
\begin{tabular}{|c|c|c|c|c|c|}
\hline 
No. & Event Name & Date & Station & ID & Comps.\tabularnewline
\hline 
\hline 
1 & Northridge, CA & Jan. 17, 1994 & Jensen Plant & 982 & 022, 292\tabularnewline
\hline 
2 & Kobe, Japan & Jan. 16, 1995 & KJMA & 1106 & 000, 090\tabularnewline
\hline 
3 & Christchurch, NZ & Feb. 21, 2011 & Hulverstone Dr. & 8090 & N04W, S86W\tabularnewline
\hline 
\end{tabular}
\end{table}

\begin{table}
\caption{\label{tab:Event-characteristics-and}Event characteristics and site
conditions of the records.}

\centering{}%
\begin{tabular}{|c|c|c|c|c|}
\hline 
No. & M\textsubscript{w} & Hyp. Depth (km) & Epi. Distance (km) & Vs30 (m/s)\tabularnewline
\hline 
\hline 
1 & 6.69 & 17.5 & 12.97 & 373\tabularnewline
\hline 
2 & 6.90 & 17.9 & 18.27 & 312\tabularnewline
\hline 
3 & 6.20 & 9.8 & 7.72 & 206\tabularnewline
\hline 
\end{tabular}
\end{table}

Table \ref{tab:Ground-motion-characteristics} lists the peak ground
acceleration (PGA), the peak ground velocity (PGV), the peak ground
displacement (PGD), as well as the 5\% pseudo-spectral acceleration
values at $T=0.2$ s and $1.0$ s for the geometric mean of the spectra
of the two horizontal components of the time series at each station.
The ratio of peak ground velocity to peak ground acceleration ($\nicefrac{\mathrm{PGV}}{\mathrm{PGA}}$)
is traditionally viewed as a measure of relative frequency content
and bandwidth of ground motion response spectrum; its multipliers
(ranging from approximately 4.0 to 5.0) are used to express the characteristic
period, $T_{s}$, that identifies the transition from constant acceleration
to constant velocity segments of the spectrum (see Fig. \ref{fig:A-two-period-seismic}).
It is noted that $T_{s}$ ``roughly corresponds to the period at
which the largest energy is imparted to the structure'' \cite{Fajfar-1}.
In this example, $\nicefrac{\mathrm{PGV}}{\mathrm{PGA}}$ varies from
0.0335 to 0.1214, or $0.13\,\mathrm{s}\lesssim T_{s}\lesssim0.61\,\mathrm{s}.$ 

\begin{table}
\caption{\label{tab:Ground-motion-characteristics}Peak ground motion and spectral
ordinates of the excitation in Example 4.}

\centering{}%
\begin{tabular}{|c|c|c|c|c|c|}
\hline 
No. & PGA (g) & PGV (cm/s) & PGD (cm) & $S_{a}\left(T=0.2\right)$ (g) & $S_{a}\left(T=1.0\right)$ (g)\tabularnewline
\hline 
\hline 
1 & 0.3465 & 31.254 & 7.487 & 1.1305 & 0.4003\tabularnewline
\hline 
2 & 0.3389 & 40.375 & 14.434 & 1.0369 & 0.5589\tabularnewline
\hline 
3 & 1.0741 & 35.342 & 9.485 & 0.9967 & 0.3699\tabularnewline
\hline 
\end{tabular}
\end{table}

The time series of the two horizontal components of ground motion
records and their Fourier amplitude spectra (see Fig. \ref{fig:Input-motion-and})
show typical nonstationarities in amplitude and frequency content
observed in records of earthquake ground motion. 

\begin{figure}
\begin{centering}
\includegraphics[scale=0.7]{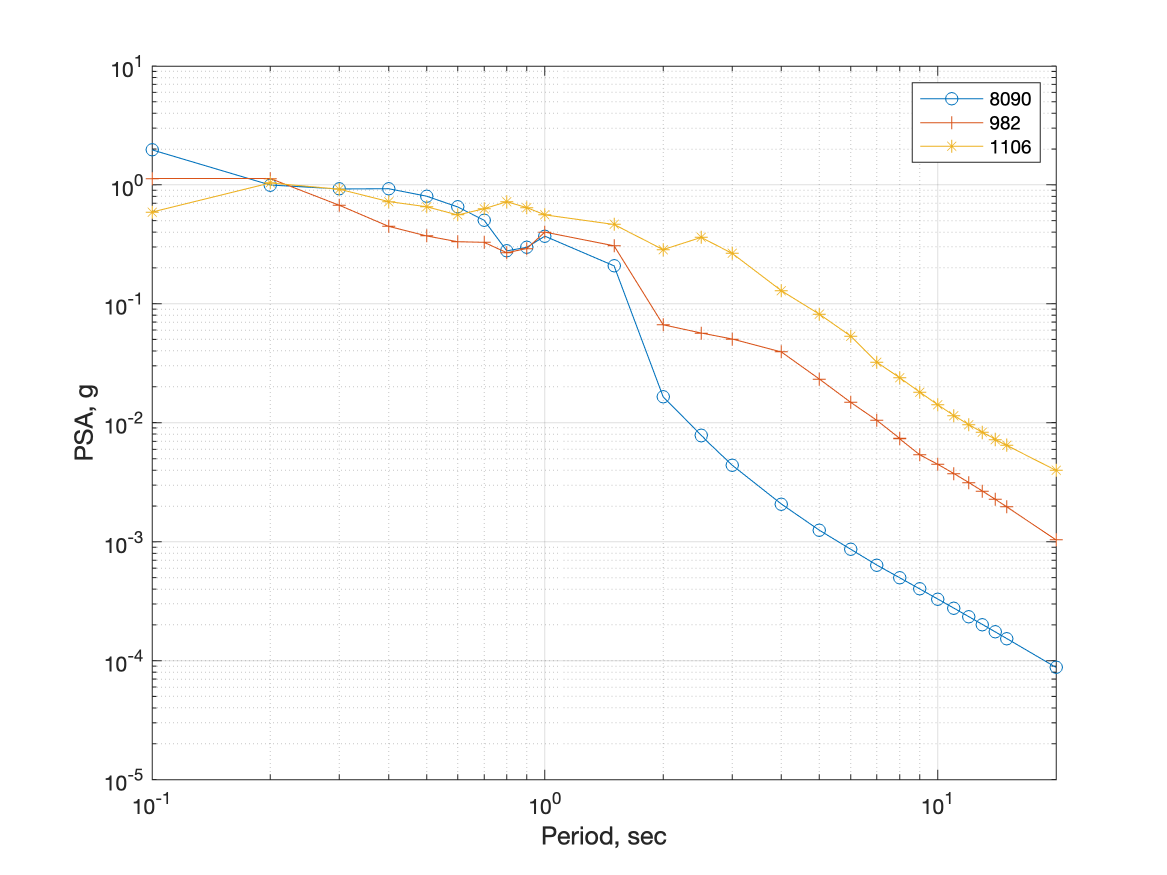}
\par\end{centering}
\caption{\label{fig:Pseudo-acceleration-spectra-of}Pseudo-acceleration spectra
of ground motions.}
\end{figure}

The geometric mean of the pseudo-acceleration response spectra of
the two horizontal components of the time series at each station is
plotted in Fig. \ref{fig:Pseudo-acceleration-spectra-of}. Note the
range of spectral ordinates in this figure from those used in Example
2 of §\ref{subsec:Example-2}. 

\begin{figure}
\begin{centering}
\includegraphics[scale=0.7]{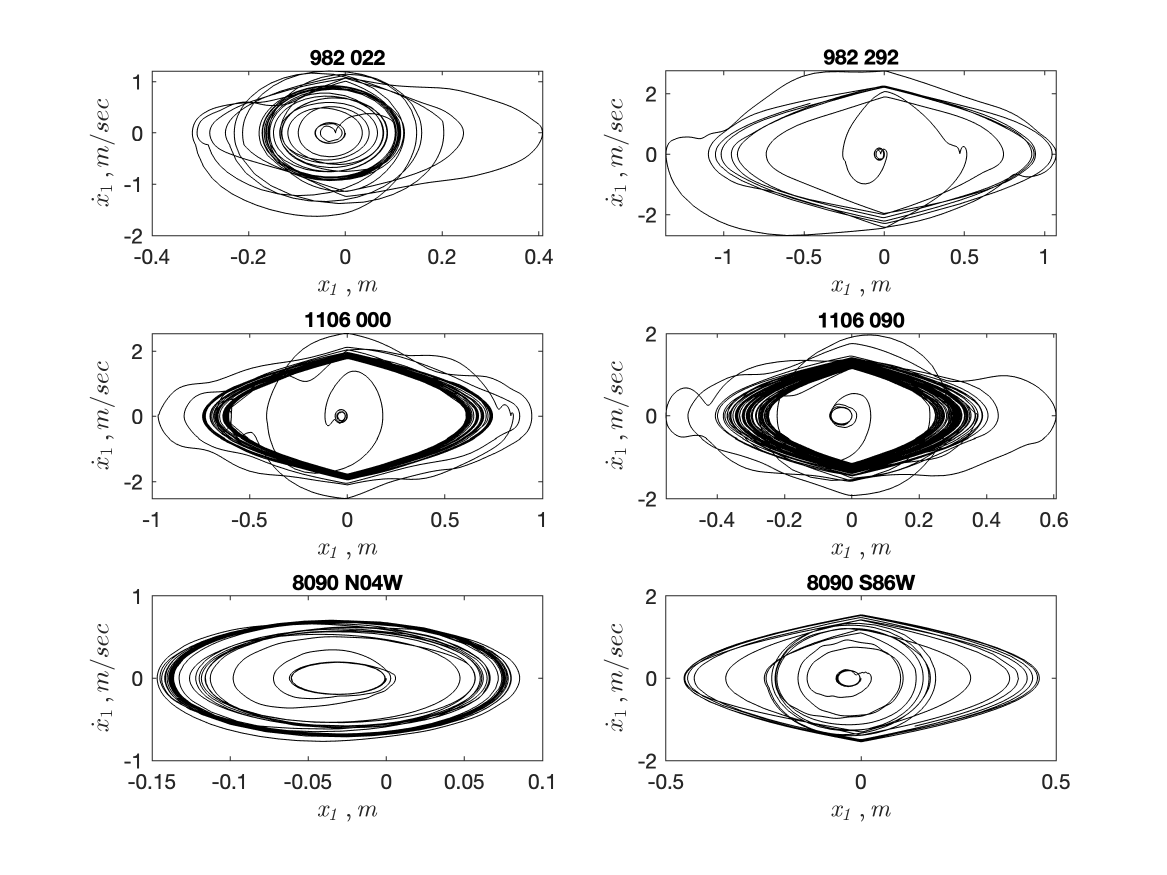}
\par\end{centering}
\caption{\label{fig:Phase-portraits-of}Phase portraits of response. The forcing
functions corresponding to the legends in Fig. \ref{fig:Input-motion-and}
are shown above each panel.}
\end{figure}

\begin{figure}
\begin{centering}
\includegraphics[scale=0.7]{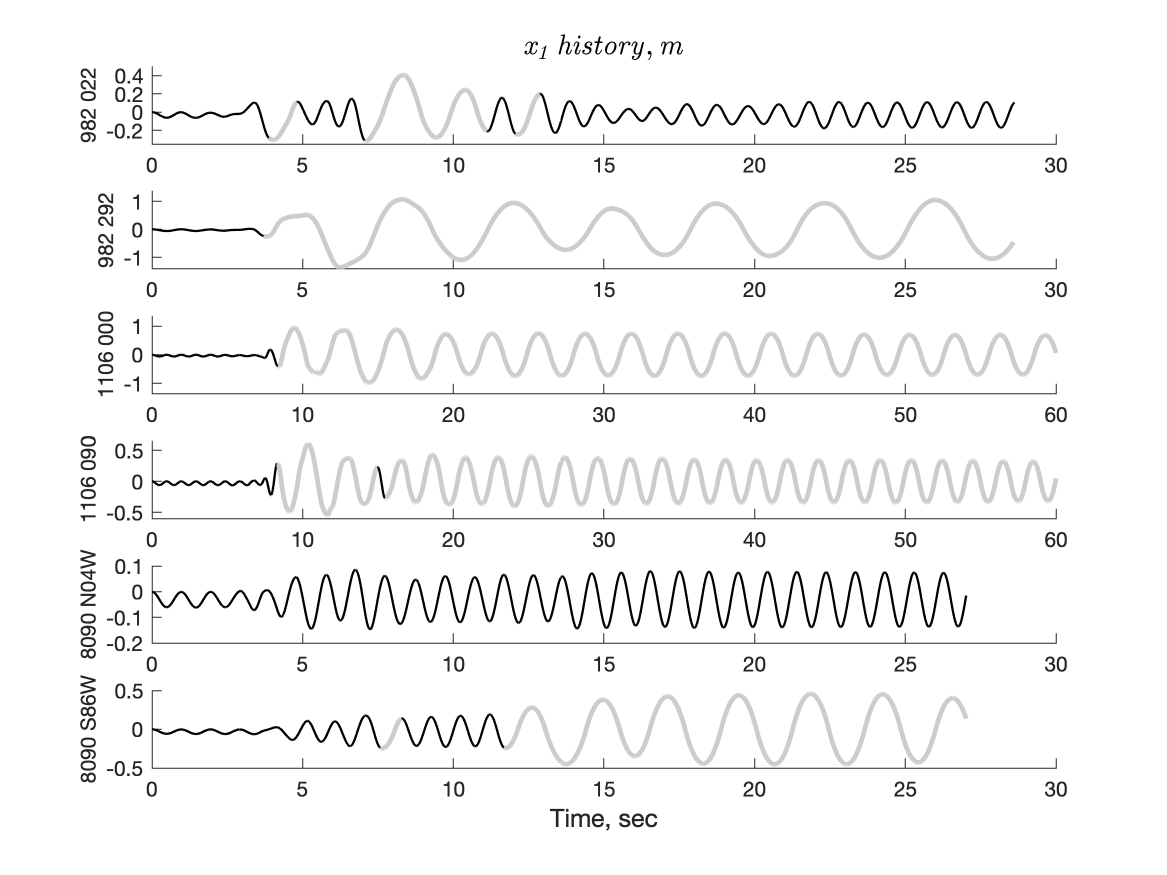}
\par\end{centering}
\caption{\label{fig:Time-history-of}Time history of displacements. The darker
thin lines show the unstepped phase and the lighter thick lines represent
the stepped phases of response. The forcing functions corresponding
to the legends in Fig. \ref{fig:Input-motion-and} are shown next
to each trace.}
\end{figure}

The solutions of \eqref{eq:4.4} and \eqref{eq:4.5} to the excitations
shown in Fig. \ref{fig:Input-motion-and} are calculated numerically
in an implicit algorithm \cite{Matlab} and are shown in Figs. \ref{fig:Phase-portraits-of}
and \ref{fig:Time-history-of}. 

In \eqref{eq:4.4}, \eqref{eq:4.5}, and \eqref{eq:4.8}:
\begin{itemize}
\item $K_{a}=1.159\times10^{9}\:\mathrm{N.m^{-1}}$,
\item $K_{b}=5.021\times10^{6}\:\mathrm{N.m^{-1}}$,
\item $\omega_{1}=6.4581$ rad/s, $\omega_{3}=0.0$ rad/s,
\item $P_{1}=1.0901$,
\item $\xi_{1}=0.05$,
\item $\omega_{2}=0.6445$ rad/s, $\omega_{4}=0.0$ rad/s,
\item $P_{2}=1.2680$, $\gamma_{1}=0.1387$, and
\item $\xi_{2}=0.0$.
\end{itemize}
The problem is stiff because of the rapid change in the trajectories
of the solution during touchdowns (slow response with rapidly changing
nearby solutions). Matlab's \emph{ode23s} and \emph{ode15s} variable-step,
variable-order numerical differentiation formula have been used with
success in this example \cite{Matlab}. The phase portraits of response
in Fig. \ref{fig:Phase-portraits-of} show some of the characteristics
of stepping response, e.g., pinching of the trajectories at $x_{1}=0$
due to discontinuity in velocity at touchdown, stepped phase of response
enveloping the unstepped phase, and asymmetry in displacements (i.e.,
oscillations about one leg between touchdowns). 

\begin{figure}
\begin{centering}
\includegraphics[scale=0.7]{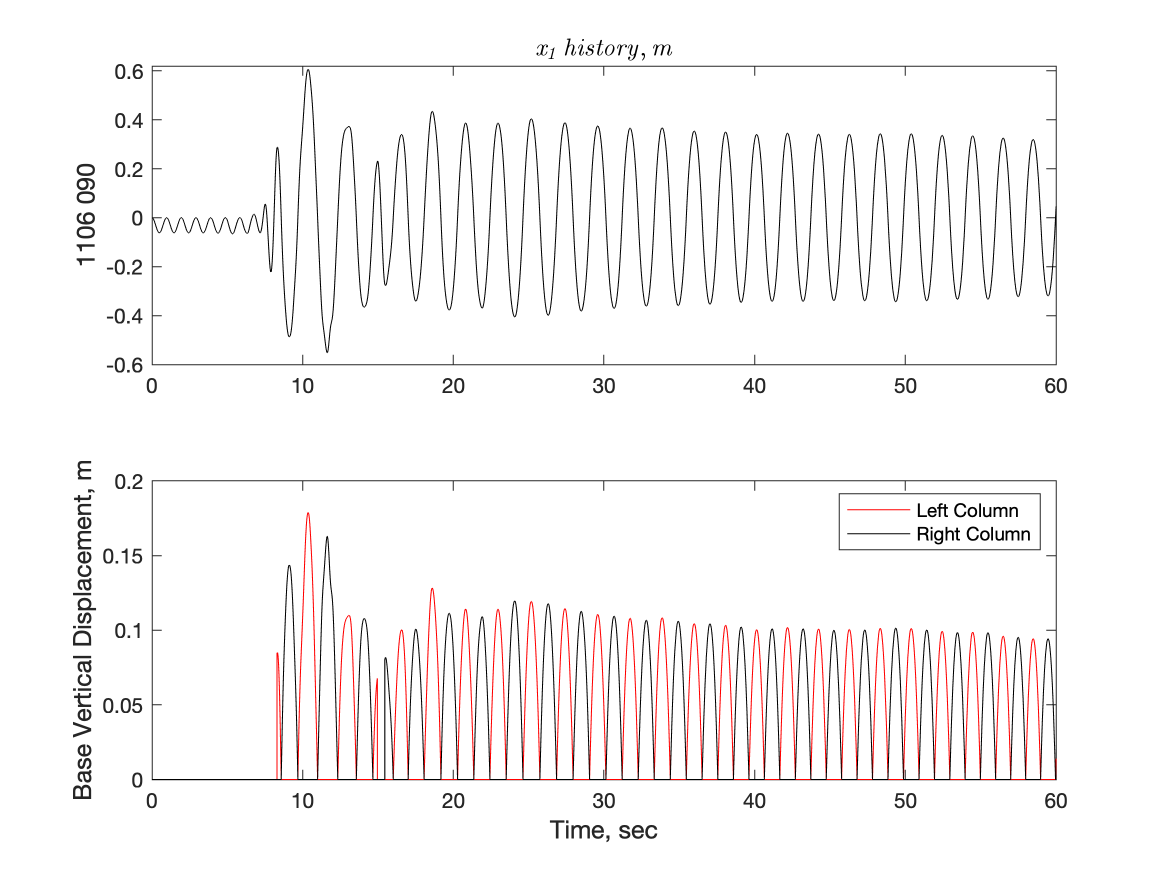}
\par\end{centering}
\caption{\label{fig:Sequence-and-amplitudes}Sequence and amplitudes of the
stepping response under \emph{1106 090}.}
\end{figure}

The time series of horizontal displacement response in Fig. \ref{fig:Time-history-of}
show that the oscillations may be shifted around one pier and also
the stiff nature of the ODEs. Notice the nonlinearity due to changes
in the oscillation periods, occasionally happening multiple times
in either directions (lengthening or shortening) within the duration
of response, which is quite unlike the response of fixed base structures
where nonlinearity in response occurs mostly in one direction due
to cumulative cyclic damage-induced stiffness softening. An expanded
view of displacement response with the sequence of stepping amplitudes
subject to forcing function \emph{1106 090 }is presented in Fig. \ref{fig:Sequence-and-amplitudes}.

The variability in displacement amplitudes shown in Fig. \ref{fig:Time-history-of}
can be taken as evidence of inadequate modeling in the iterative procedures
of AASHTO displacement-based design as they are attributed to the
violations of the aforementioned assumptions of quasi-dynamic response,
due to the temporal nonstationarity in the frequency content of the
ground motion, and other nonlinear dynamics phenomena absent in the
quasi-dynamic analysis. The similar spectral values in Fig. \ref{fig:Pseudo-acceleration-spectra-of}
produce very different dynamic response, whereas the iterative procedure
would give similar design displacements for displacement-based design. 

\section{Concluding Remarks}

Stepping (rocking) response of two-dimensional flexible frames was
studied in light of dynamic instabilities in the frameworks of equivalent
linearization of quasi-dynamic and nonlinear dynamic analysis. Closed-form
solutions and stability criteria for the displacement demand under
seismic excitations were presented. Allowing stepping response is
particularly suitable for design of tall bridges in seismic regions
because of the substantial reduction in large design force demands
at the base of the piers, leading to the possibility of reducing damage
and repair costs. Incorporation of stepping response, however, requires
establishing probabilities of reaching various levels of pier rotation
or stepping displacement for safety assessment.

The quasi-dynamic procedures of design displacement calculation in
current bridge design literature and engineering specifications, reviewed
herein, do not account for nonlinear dynamics or system and excitation
uncertainties and are only appropriate for an initial assessment of
approximate displacement demands. With sensitivity of nonlinear dynamic
response to initial conditions and implications of overturning, careful
time-stepping analyses with consideration of uncertainties in dynamic
system models, boundary conditions, and excitation is prudent in response
assessment of stepping frames. In engineered structures, for instance,
the out of plumbness is limited to $\nicefrac{\pm1}{500}^{\mathrm{th}}$
of height \cite{AISC-1} which may be taken as a guide for the range
of initial displacements that has to be examined. Additionally, the
analyses should be conducted within a large-displacement (co-rotational)
structural mechanics formulation \cite{Bathe-1,Belytschko-1} because
of the distortion of geometry during response of stepping frames. 

Care must be taken when the site and source characteristics of the
seismic excitation could cause the nonlinear oscillatory system to
experience ground motions with shifting frequencies. The selection
of ground motion time series and their processing should be performed
in such a way as to retain the temporal nonstationarities in the frequency
content of ground motion for seismic performance assessment. It was
argued that the selected accelerograms should have non-stationary
frequency content matching the expected period elongation of the structure.
A closed-form estimate of the approximate amplification of dynamic
response due to moving resonance was given and it was shown to be
a function of P- and S-wave dominant frequencies at a site. This approach
can be used to improve the equivalent linearization displacement estimate
given by the method of quasi-dynamic analysis.

\subsection{\label{sec:Future-Research-Needs}Future Research Needs}

Stepping (rocking) response is deemed an appropriate technique for
controlling seismic damage in large structures such as bridges. While
applications vary, our study was focused on planar response of two-dimensional
flexible frames. The three-dimensional and longitudinal stepping response
of framed structures, while fundamentally open to the same treatments
presented hitherto, has to consider other important phenomena such
as non-synchronous and traveling wave excitation as their effects
may contribute to unseating and propagating failure (progressive collapse)
similar to the failure of Showa Bridge during the 1964 Niigata earthquake
\cite{Showa-1}. Large out-of-phase relative displacements, induced
by foundation rocking between the pier caps, may have a similar effect
to those of liquefaction and is an important area for future work
that is largely absent in the current studies of longitudinal stepping
response of long structures.

The stepping frame problem involves both horizontal and vertical degrees
of freedom. The vertical motion may be excited because of large displacements
(deviations from tangent to the displacement curvature), pier touchdown,
or vertical ground motions. However, in the absence of significant
vertical excitation and when lateral displacements are the primary
response, the influence of vertical motion on the lateral displacements
may be ignored. For rigid blocks using large ensembles of ground motions,
the influence of the vertical component of ground motion on rocking
response has been shown to be statistically insignificant \cite{Vamvatsikos-2}.
However, for flexible portal frames, these effects call for further
investigation (§\ref{sec:The-Equations-of}), which should include
energy exchange between lateral and vertical motions and the effects
of different damping mechanisms because of the continual strain changes
during touchdown.

Procedures of displacement-based design \cite{Priestley-3} are starting
to gain attention and they are similar to the method of quasi-dynamic
rocking response analysis with equivalent linearization of a single
degree of freedom oscillator studied in §\ref{sec:Equivalent-Linearization-of}.
Therefore, the application of the presented formulations in this paper
to displacement based design may be appropriate for numerical stability
analysis and in establishing design displacements. 

Investigation of fidelity of current methods of ground motion selection
and processing in capturing the phenomenon of moving resonance in
structural response is crucial. Since moving resonance can also be
triggered by the surface waves (in addition to S-waves), the ground
motion processing methods with longer period filtering need to be
cognizant of their influence on structural response. A cross-correlation
study of $\max\left[I_{g}\left(t\right)\right]$ with $\omega_{max}-\omega_{min}$
from evolutionary power spectral density for a set of accelerograms
may indicate the efficacy of amplitude scaling methods for forcing
function alteration in the methods of probabilistic seismic performance
assessment.

\end{document}